\documentclass[12pt]{article}

\usepackage[authoryear]{natbib}
\usepackage{graphicx}
\usepackage{amssymb}
\usepackage{amsmath}
\usepackage{amsthm}
\usepackage{times}
\usepackage{bm, color}
\usepackage{subfigure}
\usepackage{multirow}
\usepackage{hyperref}
\usepackage{booktabs}
\usepackage{array}

\newcommand{\dd}{{\rm d}}
\newcommand{\Reais}{\mathbb{R}}
\newcommand{\diag}{\text{diag}}
\newcommand{\ind}{\mathbb{I}}

\newtheorem{definicao}{Definition} 
\newtheorem{remark}{Remark} 

\title{Robust inference in inflated beta regression}
\author{Francisco F.~Queiroz\footnote{Corresponding author: Francisco F.~Queiroz, email felipeq@ime.usp.br.}, Silvia L. P. Ferrari \\
{\small {\em Department of Statistics, University of S\~ao Paulo, Brazil}}}
\date{}

\begin{document}
\maketitle

\begin{abstract}

The inflated beta regression model is widely used for modeling continuous proportions with values at the boundaries. Maximum likelihood estimation for these models is well-known for its sensitivity to outliers, which can severely distort inference and lead to misleading conclusions. We propose robust estimators that mitigate the lack of robustness in maximum likelihood-based inference while preserving the simplicity and interpretability of the inflated beta framework. Additionally, an algorithm is introduced to select tuning constants based on the data's robustness requirements. The proposed estimators' asymptotic and robustness properties are studied, and robust Wald-type tests are developed. Simulation studies and a real data application highlight the advantages and practical effectiveness of the proposed robust estimators.

\end{abstract}

\section{Introduction}\label{Intro}

The inflated beta regression model, introduced by \cite{OspinaFerrari2012}, has become widely adopted for modeling data in the unit interval that includes values of zero or one. Examples include fatality or infection rate for a disease, plant cover defined as the relative area covered by a specific species, and the proportion of household income allocated to children's education. This model is a natural extension of the beta regression model \citep{FerrariCribariNeto2004}. In the inflated beta regression framework, the response variable is assumed to follow an inflated beta distribution \citep{OspinaFerrari2010}, which is a mixture of a degenerate random variable at zero or one (discrete part) and a beta-distributed random variable (continuous part). Point estimation via maximum likelihood in the inflated beta regression model is conducted separately for the discrete and continuous components. Therefore, it inherits the lack of robustness of likelihood-based inference for binary and beta regression.

The influence of outliers in logistic regression (or, more generally, in regression for binary data) has been studied by several authors. As discussed in \cite{Copas1988}, the concept of outlier is generally linked to both a geometric interpretation, where the observation is far from the data mass, and a probabilistic one, where an offensive value of the response variable occurs with very low probability if the model is correct. In logistic regression, there is little room for geometric discrepancies in the direction of the response variable, as all its values are either $0$ or $1$. Thus, outliers are defined probabilistically, occurring when the observed outcome is $1$ but the predicted probability is close to $0$, or when the observed outcome is $0$ but the predicted probability is close to $1$. Outliers can also occur in the covariate space as leverage points, which can be either good (when the difference between the observed outcome and the predicted probability is small) or bad (when this difference is large), with the latter having a strong influence on the maximum likelihood estimator \citep{Croux2003}. In recent decades, numerous studies have developed robust estimators for logistic regression parameters. \cite{Pregibon1982} pioneered this effort by proposing an M-estimator based on a Huber-type loss function. However, it is not Fisher-consistent and it is asymptotically biased. A Fisher-consistent and asymptotically normal version of Pregibon's estimator was presented by \cite{BiancoYohai1996}, though in some cases, this estimator may not exist. More recently, \cite{Croux2003} proposed a modification to ensure its existence under certain conditions. Other contributions include works by \cite{cantoni2001robust}, \cite{bondell2005minimum}, and \cite{GhoshBasu2016}. In this work, we focus on the minimum density power divergence estimator (MDPDE) proposed by \cite{GhoshBasu2016}, which minimizes the empirical version of the density power divergence introduced by \cite{basu1998robust}.

In the context of beta regression, many studies mention the lack of robustness of the maximum likelihood estimator (MLE), but there are few that propose robust estimators. \cite{ghosh2019robust} proposes the MDPDE, a robust estimator based on minimizing the empirical version of the density power divergence. \cite{ribeiro2023} define a robust estimator based on maximizing a reparametrized L$_q$-likelihood function  \citep{FerrariLaVecchia2012, LaVecchiaCamponovoFerrari2015}, called the surrogate maximum likelihood estimator (SMLE). Although the estimators proposed by \cite{ghosh2019robust} and \cite{ribeiro2023} perform well in various scenarios, they are not well-defined for all parameter values. In fact, \cite{ribeiro2023} show that MDPDE and SMLE may not be well-defined for unbounded beta densities. \cite{Maluf2024} proposed alternative versions of MDPDE and SMLE, namely LMDPDE and LSMLE, which are well-defined for all parameter values. The authors’ approach involves developing the estimators based on the distribution of the logit of the response variable, rather than directly on the beta distribution. We emphasize that these robust estimators are applicable in situations where the response variable takes values within the interval $(0,1)$, not including observations at the boundaries. 

This work aims to propose robust estimators for inflated beta regression models, extending the ideas of robustness previously developed for beta regression. While robust estimators such as LMDPDE and LSMLE have been proposed for beta regression, these approaches are not applicable to inflated beta regression due to the additional complexity introduced by the discrete component of the model. Therefore, the main contribution of this work is to build upon these methodologies, extending them to the inflated beta regression framework to address both the continuous and discrete components of the model. 

This work is organized as follows. Section \ref{inflatedbetaregression} provides the definition of inflated beta regression models, while Section \ref{robustestimators} introduces the proposed robust estimators. Empirical results are detailed in Section \ref{empiricalresults}, followed by a real data application, presented and discussed in Section \ref{application}. The paper concludes with final remarks and suggestions for future research directions.

\section{Inflated beta regression models} \label{inflatedbetaregression}

We say that a random variable $Y$ has a beta distribution with parameters $\mu \in (0,1)$ and $\phi>0$ if the probability density function of $Y$ is given by
\begin{align}\label{fdpbetaa}
f(y; \mu, \phi) = \dfrac{1}{B(\mu \phi, (1-\mu) \phi)} y^{\mu \phi - 1} (1-y)^{(1-\mu) \phi -1}, \quad 0 < y <1,
\end{align}
where $B(\cdot, \cdot)$ represents the beta function. We write $Y \sim \text{beta}(\mu, \phi)$ \citep{FerrariCribariNeto2004}. The zero-or-one inflated beta distribution was proposed by \cite{OspinaFerrari2010} and is constructed from a mixture of a degenerated random variable at a known value $c$ ($c=0$ or $c=1$) and a beta-distributed continuous random variable. The cumulative distribution function of the mixture is given by
\begin{align*}
\text{BI}_c(y; \vartheta, \mu, \phi) =  \vartheta \ind_{[c, \infty)}(y) + (1-\vartheta) F(y; \mu, \phi),
\end{align*}
where $\ind_A(y)$ denotes the indicator function that is equal to $1$ if $y \in A$ and $0$ otherwise, and $F(y; \mu, \phi)$ is the cumulative distribution function of a beta distribution with parameters $\mu$ and $\phi$. We write $\mathcal{Y} \sim \text{BI}_c(\vartheta, \mu, \phi)$. The probability density function of  $\mathcal{Y}$ is given by\footnote{The probability measure $P$ correspondent to $\text{BI}^{(c)}(y;\cdot)$ defined under the mensurable space $((0,1) \cup \{c\}, \mathcal{B})$, where $\mathcal{B}$ is the class of all Borelian subsets of $(0,1) \cup \{c\}$, is such that $P \ll \varsigma + \delta_c$, where $\varsigma$ is the Lebesgue measure and $\delta_c$ is the point of mass in $c$; that is, $\delta_c(A) = 1$, if $c \in A$ and $\delta_c(A) = 0$, if $c \notin A$, $A \in \mathcal{B}$.}
\begin{equation} \label{betai}
\text{bi}_c(y;\vartheta, \mu, \phi)  = \left\{
\begin{array}{rcl}
\vartheta,&  y=c,\\
(1-\vartheta)f(y; \mu, \phi), & y \in (0,1),\\
\end{array}
\right.
\end{equation}
where $0<\vartheta<1$, $\mu \in (0,1)$, $\phi>0$, and $f(y; \mu, \phi)$ is given by \eqref{fdpbetaa}. Note that $\vartheta= \mathbb{P}(\mathcal{Y}=c)$ represents the probability of observing zero or one (depending on the value of $c$) and $\mu$ and $\phi$ are the parameters of the beta distribution and represent the conditional mean and precision of $\mathcal{Y}$ given that $\mathcal{Y}\in (0,1)$. For further details, see \cite{OspinaFerrari2010}.

The class of zero-or-one inflated beta regression models is defined as follows. Let $\mathcal{Y}_1, \ldots, \mathcal{Y}_n$ independent random variables, where $\mathcal{Y}_i \sim \mbox{BI}_c ( \vartheta_i, \mu_i, \phi_i)$, for $i=1, \ldots, n$, and $c=0$ or $c=1$. We assume that
\begin{align}\label{ligacaoBETAINF}
g_\vartheta (\vartheta_i) =  \bm{\mathcal{S}}_i^{\top} \bm{\kappa}, \quad 
g_\mu (\mu_i)= \bm{\mathcal{X}}_i^{\top} \bm{\beta}, \quad 
g_\phi (\phi_i)= \bm{\mathcal{Z}}_i^{\top} \bm{\gamma},
\end{align}
where $\bm{\kappa}=(\kappa_1, \ldots, \kappa_{p_0})^{\top} \in \Reais^{p_0}$, $\bm{\beta}=(\beta_1, \ldots, \beta_{p_1})^{\top} \in \Reais^{p_1}$ and $\bm{\gamma}=(\gamma_1, \ldots, \gamma_{p_2})^{\top} \in \Reais^{p_2}$ are the unknown parameters vectors, $\bm{\mathcal{S}}_i = (\mathcal{S}_{i1}, \ldots, \mathcal{S}_{i{p_0}})^{\top}$, $\bm{\mathcal{X}}_i = (\mathcal{X}_{i1}, \ldots, \mathcal{X}_{i{p_1}})^{\top}$ and $\bm{\mathcal{Z}}_i = (\mathcal{Z}_{i1}, \ldots, \mathcal{Z}_{i{p_2}})^{\top}$
are the observations of the covariates with ${p_0}+{p_1}+{p_2}<n$. We assume that the link functions $g_\vartheta:(0,1)\rightarrow \Reais$, $g_\mu:(0,1)\rightarrow \Reais$ and $g_\phi:(0,\infty)\rightarrow \Reais$ are strictly monotonic and twice differentiable. A common choice for $g_\vartheta$ and $g_\mu$ is the logit function and for $g_\phi$, the logarithm function. The inflated beta regression parameters are interpreted in terms of the $\vartheta_i$, $\mu_i$ and $\phi_i$, which represent, respectively, $\mathbb{P}(\mathcal{Y}_i = c)$ and the mean and the precision of the conditional distribution of $\mathcal{Y}_i$ given that $\mathcal{Y}_i \in (0,1)$. 

Let $\bm{\upsilon}=(\bm{\kappa}^{\top}, \bm{\beta}^{\top},\bm{\gamma}^{\top})^{\top}$ the parameter vector. The likelihood function of $\bm{\upsilon}$ is given by
\[
L(\bm{\upsilon}) = \prod_{i=1}^{n} \text{bi}_c(y_i;\vartheta_i, \mu_i, \phi_i) = L_1(\bm{\kappa}) L_2(\bm{\beta}, \bm{\gamma}),
\]
where
\[
L_1(\bm{\kappa}) = \prod_{i=1}^{n} \vartheta_i^{\ind_{\{c\}}(y_i)} (1-\vartheta_i)^{1-\ind_{\{c\}}(y_i)}, \quad 
L_2(\bm{\beta}, \bm{\gamma}) =  \prod_{i: y_i \in (0,1)} f(y_i;\mu_i, \phi_i).
\]

Note that $L(\bm{\upsilon})$ is factorized in two terms: one that depends only on $\bm{\kappa}$ and another that depends only on $( \bm{\beta}^{\top},\bm{\gamma}^{\top})^{\top}$. Thus, we say that the parameter vector is separable and the maximum likelihood inference may be carried out separately for $\bm{\kappa}$ (discrete part) and $( \bm{\beta}^{\top},\bm{\gamma}^{\top})^{\top}$ (continuous part). In addition, $L_1(\bm{\kappa})$ is the likelihood function of a generalized linear model for a binary response \citep{McCullaghNelder1989} and $L_2(\bm{\beta}, \bm{\gamma})$ is the likelihood function of a beta regression model where the response variable is restricted to the observations on $(0,1)$. 

The log-likelihood function of $\bm{\upsilon}$ is given by
\begin{equation}\label{logverBETAinf}
\ell (\bm{\upsilon}) = \ell_1(\bm{\kappa}) + \ell_2(\bm{\beta}, \bm{\gamma}), 
\end{equation}
where
\[
 \ell_1(\bm{\kappa}) = \sum_{i=1}^{n} \ell_{1i}(\vartheta_i), \quad
 \ell_2(\bm{\beta}, \bm{\gamma}) = \sum_{i: y_i \in (0,1)}\ell_{2i} (\mu_i,\phi_i),
\]
in which 
\begin{align*} 
\begin{split}
\ell_{1i}(\vartheta_i) & = \ind_{\{c\}} (y_i) \log (\vartheta_i) + [1-\ind_{\{c\}} (y_i)] \log (1-\vartheta_i), \\
\ell_{2i} (\mu_i,\phi_i) & = \log \Gamma (\phi_i) - \log \Gamma (\mu_i \phi_i) - \log \Gamma ((1-\mu_i)\phi_i) + (\mu_i \phi_i - 1) \log( y_i/(1-y_i)) \\
& + (\phi_i - 2) \log(1-y_i).
\end{split}
\end{align*}
Observe that, for $i=1, \ldots, n$, the random variable $\ind_{\{c\}} (\mathcal{Y}_i)$ is Bernoulli-distributed with $\mathbb{P}(\ind_{\{c\}} (\mathcal{Y}_i)=1) =  \vartheta_i$. The MLE of $\bm{\upsilon}$ is obtained by maximizing \eqref{logverBETAinf}.

Let
\begin{align}\label{conjP}
\wp = \wp\left(\ind_{\{c\}} (\mathcal{Y}_1), \ldots, \ind_{\{c\}} (\mathcal{Y}_n)\right) =  \left\{i \in \{1, \ldots, n\}: \ind_{\{c\}} (\mathcal{Y}_i) = 0\right\},
\end{align}
of length $n^\dagger$, representing the set of indices of the observations on $(0,1)$. In the following, we use $\wp$ to denote both the random set $\wp\left(\ind_{\{c\}} (\mathcal{Y}_1), \ldots, \ind_{\{c\}} (\mathcal{Y}_n)\right)$ or the observed set $\wp\left(\ind_{\{c\}} (y_1), \ldots, \ind_{\{c\}} (y_n)\right)$, depending on the context. 

Let
\begin{equation*}
\mathcal{Y}_i^\star = \begin{cases}
\log\left( \dfrac{\mathcal{Y}_i}{1-\mathcal{Y}_i} \right), & i \in \wp,\\
0, & i \not\in \wp,
\end{cases} \quad \quad
\mathcal{Y}_i^\dagger = \begin{cases}
\log(1-\mathcal{Y}_i), & i \in \wp,\\
0, & i \not\in \wp,
\end{cases} 
\end{equation*}
 $$\mu_i^\star = \mathbb{E}(\mathcal{Y}_i^\star | \ind_{\{c\}} (\mathcal{Y}_i) = 0)  =  \psi(\mu_{i}\phi_i)-\psi((1-\mu_{i})\phi_i),$$ and $$\mu_i^\dagger = \mathbb{E}(\mathcal{Y}_i^\dagger | \ind_{\{c\}} (\mathcal{Y}_i) = 0) =  \psi((1-\mu_{i})\phi_i)-\psi(\phi_i).$$ 
The observed values of $\mathcal{Y}_i^\star$ and $\mathcal{Y}_i^\dagger$ are denoted by $y_i^\star$ and $y_i^\dagger$, respectively. Thus, the estimating equations associated of the MLE can be written as
\begin{align}\label{eqEstimacaoBETAINF}
\begin{split}
\left\{\begin{array}{lr}
        \displaystyle \sum_{i=1}^n \textbf{U}(y_i; \bm{\kappa}) & = \bm{0}\\
 \displaystyle \sum_{i \in \wp} \textbf{U}(y_i; \bm{\beta}, \bm{\gamma}) & = \bm{0}\\
        \end{array}\right.
\end{split}
\end{align}
in which
\[
\textbf{U}(y_i; \bm{\kappa}) = \dfrac{\ind_{\{c\}} (y_i) - \vartheta_i}{\vartheta_i(1-\vartheta_i)} \dfrac{1}{g'_\vartheta(\vartheta_i)} \bm{\mathcal{S}}_i, \quad i=1, \ldots, n
\]
and
\[
\textbf{U}(y_i; \bm{\beta}, \bm{\gamma}) = \left( \phi_i \dfrac{(y_i^\star - \mu_i^\star)}{g'_\mu (\mu_i)}\bm{\mathcal{X}}_i^\top, \quad \dfrac{\mu_i(y_i^\star - \mu_i^\star) + (y_i^\dagger - \mu_i^\dagger)}{g'_\phi (\phi_i)} \bm{\mathcal{Z}}_i^\top   \right)^\top, \quad i \in \wp.
\]
 
The system of equations \eqref{eqEstimacaoBETAINF} can be solved separately, with the first equation for estimating the parameters associated with the discrete part and the second one for the continuous part. 
\begin{remark}\label{remark1}
The equation related to the discrete part is that of a regression model for binary data, where the dependent variable is $\ind_{\{c\}} (\mathcal{Y}_i) \sim \text{Bernoulli}(\vartheta_i)$, for $i=1, \ldots, n$, and the covariates $\bm{\mathcal{S}}_i$ are related to the parameter vector $\bm{\kappa}$ through the link function $g_\vartheta(\vartheta_i)$, as in \eqref{ligacaoBETAINF}.
\end{remark}

\begin{remark}\label{remark2}
 The equation for the continuous part is that of a beta regression model, where the response variable is $\mathcal{Y}_i \sim \text{beta}(\mu_i, \phi_i)$, for $i \in \wp$, and the covariates $\bm{\mathcal{X}}_i$ and $\bm{\mathcal{Z}}_i$ are related to the parameter vectors $\bm{\beta}$ and $\bm{\gamma}$ through the link functions $g_\mu(\mu_i)$ and $g_\phi(\phi_i)$, respectively; see \eqref{ligacaoBETAINF}.
\end{remark}
  Therefore, in the presence of outlier observations, maximum likelihood inference in inflated beta regression models inherits the lack of robustness observed in binary data regression models and beta regression models.

\section{Robust estimation in inflated beta regression models}\label{robustestimators}

\cite{GhoshBasu2016} developed a robust estimator, called the minimum density power divergence estimator in the context of generalized linear models. The MDPDE is obtained by minimizing the empirical version of the power divergence between two densities, which was developed by \cite{basu1998robust}. The family of power divergences of $f_{\bm{\delta}}$ with respect to $g$ is defined by
\[
d_\alpha(f_{\bm{\delta}}, g) = \displaystyle\int f_{\bm{\delta}}(v)^{1+\alpha} \dd v - \dfrac{1 + \alpha}{\alpha} \displaystyle\int f_{\bm{\delta}}(v)^{\alpha} g(v) \dd v +\dfrac{1}{\alpha} \displaystyle\int g(v)^{1+\alpha} \dd v,
\]
where $\alpha>0$ is a tuning constant and 
\[
d_0(f_{\bm{\delta}}, g) = \lim_{\alpha \rightarrow 0} d_\alpha(f_{\bm{\delta}}, g)  =\displaystyle\int  \log \left( \dfrac{g(v)}{f_{\bm{\delta}}(v)} \right) g(v) \dd v.
\]

In practice, $g$ represents the density of the data and $f_{\bm{\delta}}$ the density of the model (which depends on unknown parameters). Densities can be replaced by probability functions, in the context of discrete random variables; in this case, integrals are replaced by summations. Finally, MDPDE is obtained by minimizing the empirical version of $d_\alpha(f_{\bm{\delta}}, g)$ which, up to a term that does not depend on $\bm{\delta}$, is given by
\begin{align*}
\mathcal{H}_n(\bm{\delta}) = \dfrac{1}{n} \displaystyle\sum_{i=1}^n \mathcal{V}_i(v_i; \bm{\delta}),
\end{align*}
where
\[
\mathcal{V}_i(v_i; \bm{\delta}) = \displaystyle\int f_{\bm{\delta}}(v)^{1+\alpha} \dd v - \dfrac{1 + \alpha}{\alpha} f_{\bm{\delta}}(v_i)^{\alpha},
\]
provided that the integral converges. Note that if $\alpha=0$, the power divergence reduces to the Kullback-Leibler divergence, and MDPDE coincides with MLE. Additionally, for $\alpha\in (0,1)$ the family of power divergences provides a smooth path between the Kullback-Leibler divergence and the $L_2$ distance between densities \citep{basu1998robust}. We consider $0 \leq \alpha<1$.

We now consider a robust estimation of $\bm{\kappa}$. Let $Y_i^c = \ind_{\{c\}} (\mathcal{Y}_i)$, for $i=1, \ldots, n$, which are independent Bernoulli-distributed random variables with parameter $\vartheta_i = g_\vartheta^{-1}(\bm{\mathcal{S}_i}^{\top} \bm{\kappa})$; see Remark \ref{remark1}. The MDPDE for $\bm{\kappa}$ is obtained by minimizing
\begin{align*}
\mathcal{H}_n(\bm{\kappa}) = \dfrac{1}{n} \displaystyle\sum_{i=1}^n \mathcal{V}_i(y_i^c; \bm{\kappa}),
\end{align*}
where
\[
\mathcal{V}_i(y_i^c; \bm{\kappa}) = \vartheta_i^{1+\alpha} + (1-\vartheta_i)^{1+\alpha} - \dfrac{1 + \alpha}{\alpha} f_{\bm{\kappa}}(y_i^c; \vartheta_i)^{\alpha}, \] with $f_{\bm{\kappa}}(y_i^c; \vartheta_i) = \vartheta_i^{y_i^c}(1-\vartheta_i)^{1- y_i^c}$ being the probability function of $Y_i^c$. Thus, the estimating equation is given by
\begin{align}\label{eqestimacaologistica}
\displaystyle\sum_{i=1}^n \textbf{U}^*(y_i^c; \bm{\kappa}) = \bm{0},
\end{align}
where
\[
\textbf{U}^*(y_i^c; \bm{\kappa}) = \dfrac{(1+\alpha)}{\vartheta_i (1-\vartheta_i)} \left[ (1-\vartheta_i)\vartheta_i^{1+\alpha} - \vartheta_i(1-\vartheta_i)^{1+\alpha} - (y_i^c - \vartheta_i)f_{\bm{\kappa}}(y_i^c; \vartheta_i)^{\alpha} \right] \dfrac{1}{g'_\vartheta(\vartheta_i)} \bm{\mathcal{S}}_i.
\]

Note that 
\begin{align} \label{umbiesednessofUk}
 \mathbb{E}[ (Y_i^c - \vartheta_i)f_{\bm{\kappa}}(Y_i^c; \vartheta_i)^{\alpha} ] = (1-\vartheta_i)\vartheta_i^{1+\alpha} - \vartheta_i(1-\vartheta_i)^{1+\alpha} , \quad \forall \bm{\upsilon} \in \Reais^{p_0+p_1+p_2}, 
 \end{align}
which implies that $\mathbb{E}[\textbf{U}^*(Y_i^c; \bm{\kappa})] = \bm{0}$, $\forall \bm{\upsilon} \in \Reais^{p_0+p_1+p_2}$. That is, the estimating function in \eqref{eqestimacaologistica} is unbiased. This result will be used to show the Fisher-consistency of the proposed estimators for the inflated beta regression models.

If $\alpha$ is not very close to zero, observations with large values of $(y_i^c - \vartheta_i)$ receive small weights, given by $f_{\bm{\kappa}}(y_i^c; \vartheta_i)^{\alpha}$. Thus, bad leverage points, which have a large influence on MLE, have small influence on MDPDE. Consequently, the weight mitigates the effect of the bad leverage points. The tuning constant $\alpha$ provides a balance between efficiency ($\alpha=0$ leads to MLE) and robustness.
%
%

In what follows, we will address robust estimation of the parameters associated with the continuous part of the inflated beta regression model. From now on, we assume that the set $\wp$ is fixed. Let $y_i$, for $i \in \wp$, be the observed data from a beta regression model defined by \eqref{fdpbetaa} and by $g_\mu (\mu_i)$ and $g_\phi (\phi_i)$ in \eqref{ligacaoBETAINF}. From Remark \ref{remark2}, these data are used to estimate $\bm{\theta} = (\bm{\beta}^\top, \bm{\gamma}^\top)^\top$.

We propose two robust estimators for $\bm{\theta}$, which are based on the maximization of a reparametrized L$_q$-likelihood and on the minimization of an empirical version of a density power divergence. For beta regression models, these estimators, called SMLE and MDPDE, were initially proposed by \cite{ribeiro2023} and \cite{ghosh2019robust}, respectively. However, \cite{ribeiro2023} noted that they depend on specific restrictions on the parameter space to be well-defined. \cite{Maluf2024} showed that the main limitation of the SMLE and the MDPDE comes from the fact that the beta densities are not closed under power transformations. Given a density $v$ and a constant $\xi\in(0,\infty)$, the power transformation is
\begin{equation*}
 v^{\left(\xi\right)}(y)=\frac{v(y)^{\xi}}{\int v(y)^{\xi} \dd y} \varpropto v(y)^\xi, \quad \forall y \text{~in the support,}
\end{equation*}
provided that $\int v(y)^{\xi} \dd y<\infty$. For the beta density \eqref{fdpbetaa},
\[
f(y; \mu, \phi)^\xi \varpropto y^{\xi (\mu \phi -1) } (1-y)^{\xi [(1-\mu) \phi -1] },
\]
which is integrable for all $\xi \in (0, \infty)$ if and only if $\mu \phi \geq 1$ and $(1-\mu) \phi\geq 1$, that is, if and only if the beta density \eqref{fdpbetaa} is bounded.

\cite{Maluf2024} then proposed robust estimators that are well-defined and preserve robustness properties even for unbounded beta densities. They followed the ideas of \cite{ribeiro2023} and \cite{ghosh2019robust}, but considering the logit transformation $Y^\star = \log(Y/(1-Y))$. If  $Y \sim \text{beta}(\mu, \phi)$, the density function of $Y^\star = \log(Y/(1-Y))$ is given by
\begin{align}\label{egbdef}
h(y^\star;\mu, \phi) = \dfrac{1}{B(\mu \phi, (1-\mu) \phi)} \dfrac{e^{-y^\star(1-\mu) \phi}}{(1+e^{-y^\star})^\phi}, \quad y^\star \in \Reais,
\end{align}
where $0<\mu <1$ and $\phi>0$. Note that $h(y^\star;\mu,\phi)^\xi \propto h(y^\star;\mu,\xi \phi)$, for all $y^\star\in\mathbb{R}$, $\mu \in (0,1)$, and $\phi, \xi >0$. That is, the class of the densities in \eqref{egbdef} is closed under power transformations. \cite{Maluf2024} called the new estimators logit SMLE (LSMLE) and logit MDPDE (LMDPDE).

The L$_q$-likelihood function \citep{FerrariYang2010} of a parameter vector $\bm{\delta}$ based on a density function $f_{\bm{\delta}}(v)$ is given by
\[
\ell_q(\bm{\delta}) = \displaystyle \sum_{i} L_q(f_{\bm{\delta}}(v_i)),
\]
where 
\begin{equation*}
L_q(u) = \begin{cases}
(u^{1-q} -1)/(1-q), & q \not= 1,\\
\log(u), & q = 1,
\end{cases} 
\end{equation*}
$0 < q \leq 1$ is a tuning constant. We set $\alpha = 1 - q$ to match the previous notation\footnote{Note that this tuning constant does not need to be equal to that of the robust estimator of $\bm{\kappa}$, the parameter vector of the discrete part.}. The maximum L$_q$-likelihood estimator (ML$_q$E) is obtained maximizing $\ell_{1-\alpha}(\bm{\theta})$. If $\alpha = 0$, ML$_q$E coincides with  MLE. The estimating equation associated with ML$_q$E is 
\begin{align}\label{eqEstimacaoBETALq}
\begin{split}
\displaystyle \sum_{i} \textbf{U}(v_i; \bm{\delta})f_{\bm{\delta}}(v_i)^\alpha & = \bm{0},
\end{split}
\end{align}
where $\textbf{U}(v_i; \bm{\delta})$ is the first derivative of $\log f_{\bm{\delta}}(v_i)$ with respect to $\bm{\delta}$.  The estimating function is not guaranteed to be unbiased unless $\alpha =0$, so the ML$_q$E may not be Fisher-consistent. A Fisher-consistent estimator, named surrogate maximum likelihood estimator, can be obtained by maximizing a reparameterized L$_q$-likelihood, provided that the model density is closed under power transformations  \citep{FerrariLaVecchia2012, LaVecchiaCamponovoFerrari2015}. The SMLE maximizes the L$_q$-likelihood in the parametrization $\tau_{1/(1-\alpha)}(\bm{\delta})$, where $\tau_\omega(\bm{\delta}): \bm{\Theta} \longmapsto \bm{\Theta}$ is a continuous function satisfying $f_{\tau_\omega(\bm{\delta})}(v) = f_{\bm{\delta}}^{(\omega)}(v)$ for all $v$ in the support and $\bm{\Theta}$ is the parameter space of $\bm{\delta}$, given by 
\[
\ell_{1-\alpha}^*(\bm{\delta}) = \displaystyle \sum_{i} L_{1-\alpha}\left(f_{\bm{\delta}}^{\left(\frac{1}{1-\alpha}\right)}(v_i)\right).
\]

The LSMLE for $\bm{\theta}$  is obtained by maximizing
\[
\ell_{1-\alpha}^*(\bm{\theta}) = \displaystyle \sum_{i \in \wp} L_{1-\alpha}\left(h_{\bm{\theta}}^{\left(\frac{1}{1-\alpha}\right)}(y_i^\star; \mu_i, \phi_i)\right),
\] 
where $h_{\bm{\theta}}^{\left(\frac{1}{1-\alpha}\right)}(y_i^\star; \mu_i, \phi_i) = h_{\bm{\theta}}(y_i^\star; \mu_i, \phi_{i, (1-\alpha)^{-1}})$, where $\phi_{i,\alpha} = \phi_i \alpha$ with $0 \leq \alpha<1$. 
Note that $h_{\bm{\theta}}(y_i^\star; \mu_i, \phi_{i, (1-\alpha)^{-1}})$ is the density function of the logit transformation of a variable that follows a modified beta regression model with mean and precision submodels given respectively by
\[
g^*_\mu(\mu_i) = g_\mu(\mu_i) = \bm{\mathcal{X}}_i^{\top} \bm{\beta}, \quad \quad g^*_\phi(\phi_i) = g_\phi({\phi}_{i, 1-\alpha}) = \bm{\mathcal{Z}}_i^{\top} \bm{\gamma},
\]
which will be denoted by $h^\ast_{\bm{\theta}}(y_i^\star; \mu_{i}, \phi_{i})$. The estimating equation is given by
\begin{align}\label{eqEstimacaoEGBLqrep}
\begin{split}
\displaystyle \sum_{i \in \wp} \textbf{U}^\ast(y_i^\star; \bm{\theta})h^\ast_{\bm{\theta}}(y_i^\star; \mu_i, \phi_i)^\alpha & = \bm{0},
\end{split}
\end{align}
and $\textbf{U}^\ast(y_i^\star; \bm{\theta}) = (\partial/\partial \bm{\theta})\log h^\ast_{\bm{\theta}}(y_i^\star; \mu_i, \phi_i)$ is the modified score vector for the $i$-th observation given by
\[
\textbf{U}^\ast(y_i^\star; \bm{\theta}) = \left( \phi_{i} \dfrac{(y_i^\star - {\mu}_i^\star)}{g'_\mu (\mu_{i})}\bm{\mathcal{X}}_i^\top, \quad (1-\alpha)^{-1} \dfrac{\mu_{i}(y_i^\star - {\mu}_i^\star) + (y_i^\dagger - {\mu}_i^\dagger)}{g'_\phi ({\phi}_{i, 1-\alpha})} \bm{\mathcal{Z}}_i^\top   \right)^\top.
\]

The LMDPDE is obtained by minimizing 
\begin{align}\label{funcaoHMDPDEEGB}
\mathcal{H}_n(\bm{\theta}) = \dfrac{1}{n} \displaystyle\sum_{i \in \wp} \mathcal{V}_i(y_i^\star; \bm{\theta}),
\end{align}
where 
\[
\mathcal{V}_i(y_i^\star; \bm{\theta}) = \mathcal{K}_{i, 1+\alpha}(\bm{\theta}) - \dfrac{1 + \alpha}{\alpha} h_{\bm{\theta}}(y_i^\star; \mu_i, \phi_i)^{\alpha},
\]
and
$$\mathcal{K}_{i, 1+\alpha}(\bm{\theta}) = \displaystyle\int_{-\infty}^\infty h_{\bm{\theta}}(y^\star; \mu_i, \phi_i)^{1+\alpha} \dd y^\star = \dfrac{B(\mu_i \phi_i(1+\alpha), (1-\mu_i) \phi_i (1+\alpha))}{B(\mu_i \phi_i, (1-\mu_i) \phi_i)^{1+\alpha}},$$
for $0\leq\alpha<1$. Note that the integral is finite for all $0\leq\alpha<1$. The estimating equation is given by
\begin{align}\label{eqEstimacaoEGBMDPDE}
\displaystyle \sum_{i \in \wp} [\textbf{U}(y_i; \bm{\theta})h_{\bm{\theta}}(y_i^\star; \mu_i, \phi_i)^\alpha  - \mathcal{E}_{i,1-\alpha}(\bm{\theta})]& = \bm{0},
\end{align}
where $\mathcal{E}_{i,1-\alpha}(\bm{\theta}) = \mathbb{E}\left[\textbf{U}(\mathcal{Y}_i; \bm{\theta})h_{\bm{\theta}}(\mathcal{Y}_i^\star; \mu_i, \phi_i)^\alpha \middle| \ind_{\{c\}}(\mathcal{Y}_i) = 0 \right] = \left( \Lambda_{1, i}^{(1+\alpha)}\bm{\mathcal{X}}_i^\top, \Lambda_{2, i}^{(1+\alpha)}\bm{\mathcal{Z}}_i^\top \right)^\top$, with
\begin{align*}
\Lambda_{1, i}^{(\alpha)} = \dfrac{\phi_i \mathcal{K}_{i, \alpha}(\bm{\theta}) }{g'_\mu(\mu_i)}(\mu_{i, \alpha}^{ \star} -\mu^\star_{i} ), \quad \text{and} \quad \Lambda_{2, i}^{(\alpha)} = \dfrac{ \mathcal{K}_{i, \alpha}(\bm{\theta}) }{g'_\phi(\phi_i)}[\mu_i(\mu_{i, \alpha}^{ \star} -\mu^\star_{i}) + (\mu_{i, \alpha}^{ \dagger} -\mu^\dagger_{i})],
\end{align*}
in which $\mu_{i, \alpha}^{ \star} = \psi(\mu_{i} \phi_{i, \alpha}) - \psi((1-\mu_{i}) \phi_{i, \alpha})$ and $\mu_{i, \alpha}^{ \dagger} = \psi((1-\mu_{i}) \phi_{i, \alpha}) - \psi( \phi_{i,\alpha})$. 

In conclusion, the proposed robust estimators solve the following estimating equation
\begin{align} \label{eqestESTROB}
\textbf{U}(\bm{y}; \bm{\upsilon}) = \bm{0},
\end{align}
with $\textbf{U}(\bm{y}; \bm{\upsilon}) = (\textbf{U}(\bm{\kappa})^\top, \textbf{U}(\bm{\theta})^\top)^\top$, where $\textbf{U}(\bm{\kappa})$ is given by the left-hand side of \eqref{eqestimacaologistica} and $\textbf{U}(\bm{\theta})$ depends on the specific estimator under consideration. In the following, we formally define and name the proposed estimators.

\begin{definicao}[M-LSE] M-LSE solves \eqref{eqestESTROB} with $\textbf{U}(\bm{\theta})$ given by the left-hand side of \eqref{eqEstimacaoEGBLqrep}. This estimator considers the MDPDE for the discrete part and the LSMLE for the continuous part.
\end{definicao}

\begin{definicao}[M-LME] M-LME solves \eqref{eqestESTROB} with $\textbf{U}(\bm{\theta})$ given by the left-hand side of \eqref{eqEstimacaoEGBMDPDE}. This estimator considers the MDPDE for the discrete part and the LMDPDE for the continuous part.
\end{definicao}


The asymptotic normality of the proposed estimators arises from the fact that they are M-estimators. Therefore, under usual regularity conditions, the M-LSE and M-LME estimators are asymptotically normally distributed with mean $\bm{\upsilon}$ and variance-covariance matrix $V_\alpha(\bm{\upsilon})$ given by
\begin{align*}
V_\alpha(\bm{\upsilon}) = 
\begin{bmatrix}
V_{\bm{\kappa}, \alpha}(\bm{\kappa}) & \bm{0}\\
\bm{0} & V_{\bm{\theta}, \alpha}(\bm{\upsilon})
\end{bmatrix},
\end{align*}
where
$$V_{\bm{\kappa}, \alpha}(\bm{\kappa}) = \textbf{A}_\alpha^{-1}(\bm{\kappa}) \textbf{B}_\alpha(\bm{\kappa})\textbf{A}_\alpha^{-1}(\bm{\kappa}),$$
with
\begin{align} \label{MatrizABMDPDEdisc}
\begin{split}
\textbf{A}_\alpha(\bm{\kappa}) &= \displaystyle\sum_{i=1}^n \mathbb{E} \left[\dfrac{\partial}{\partial \bm{\kappa}^\top} \left( \dfrac{\partial}{\partial \bm{\kappa}} \mathcal{V}_i(Y_i^c; \bm{\kappa}) \right) \right] =  \bm{\mathcal{S}}^\top \textbf{M} \bm{\Lambda} \textbf{T}_{\vartheta}^2  \bm{\mathcal{S}},\\
\textbf{B}_\alpha(\bm{\kappa}) & = \displaystyle\sum_{i=1}^n \mathbb{E} \left[ \left( \dfrac{\partial}{\partial \bm{\kappa}}\mathcal{V}_i(Y_i^c; \bm{\kappa})\right) \left( \dfrac{\partial}{\partial \bm{\kappa}} \mathcal{V}_i(Y_i^c; \bm{\kappa})\right)^\top \right] =  \bm{\mathcal{S}}^\top \textbf{M}^2 \bm{\Lambda} \textbf{T}_{\vartheta}^2  \bm{\mathcal{S}},
\end{split}
\end{align}
where $\bm{\mathcal{S}} = (\bm{\mathcal{S}}_1, \ldots, \bm{\mathcal{S}}_{n})^\top$, $\textbf{M} = \diag\{ \text{M}_i; i=1, \ldots, n\}$, $\bm{\Lambda} = \diag\{ [\vartheta_i(1-\vartheta_i)]^{-1}; i=1, \ldots, n \}$, $\textbf{T}_{\vartheta} = \diag\{ [g'_\vartheta(\vartheta_i)]^{-1}; i=1, \ldots, n \}$ and $
\text{M}_i = \text{M}_i(\bm{\kappa}, \alpha) = (1+\alpha)\left[ (1-\vartheta_i)\vartheta_i^\alpha + \vartheta_i (1-\vartheta_i)^\alpha \right)]$. The detailed calculations for deriving $V_{\bm{\kappa}, \alpha}(\bm{\kappa})$ are provided in the Supplementary Material. The matrix $V_{\bm{\theta}, \alpha}(\bm{\upsilon})$ depends on the robust estimator under consideration. In the Appendix, we present the expressions for this matrix for each of the estimators under study. The derivation of this matrix is quite similar for both estimators. In the Supplementary Material, we provide detailed steps for obtaining $V_{\bm{\theta}, \alpha}(\bm{\upsilon})$ for M-LSE.

To establish the property of Fisher consistency, it suffices to demonstrate that the estimating function $\textbf{U}(\bm{y}; \bm{\upsilon})$ is unbiased, or equivalently, that both $\textbf{U}(\bm{\kappa})$ and $\textbf{U}(\bm{\theta})$ are unbiased. From \eqref{umbiesednessofUk}, it follows that $\mathbb{E}[\textbf{U}^*(Y_i^c; \bm{\kappa})] = \bm{0}$, $\forall \bm{\upsilon} \in \Reais^{p_0+p_1+p_2}$. Therefore, $\textbf{U}(\bm{\kappa})$ is unbiased. In the Supplementary Material, we show that the estimating function $\textbf{U}(\bm{\theta})$ is unbiased for each of the considered estimators.

The proposed estimators are B-robust, meaning that the maximum bias induced by infinitesimal contaminations in the data is finite. This is a desirable property for any robust estimator. Since the proposed estimators are M-estimators, demonstrating B-robustness requires showing that the components of the estimating function are bounded; specifically, that both $\textbf{U}(\bm{\kappa})$ and $\textbf{U}(\bm{\theta})$ are bounded \citep[Section 2.1]{hampel2011robust}. The estimating function $\textbf{U}(\bm{\kappa})$ is clearly bounded, since $y^c \in \{0,1\}$. The proofs that $\textbf{U}(\bm{\theta})$ for each estimator is bounded can be found in \cite{Maluf2024}.

Now, consider the problem of performing a hypothesis test for a single component of the parameter vector $\bm{\upsilon}$, say $\upsilon_j$. The null hypothesis is $\mathcal{H}_0: \upsilon_j = \upsilon_j^{(0)}$ to be tested against on the alternative $\mathcal{H}_1: \upsilon_j \not= \upsilon_j^{(0)}$. The robust Wald-type test statistic is defined by
\[
W_{j} = \left(\dfrac{ \widehat{\upsilon}_{ j} - \upsilon_j^{(0)} }{\text{se}\left( \widehat{\upsilon}_{ j} \right)}\right)^2,
\]
where $\text{se}\left( \widehat{\upsilon}_{ j} \right)$ is the asymptotic standard error of $\widehat{\upsilon}_{ j}$, and $\widehat{\upsilon}_{ j}$ represents any of the proposed robust estimators. Under the null hypothesis, $W_{j}$ has an asymptotic $\chi_1^2$ distribution. If the tuning constants are equal to zero, $W_{j}$ coincides with the usual Wald statistic.

A key aspect of robust inference is the choice of the tuning constant, which balances robustness and efficiency; higher values of the tuning constants increase robustness but reduce efficiency. Most studies recommend values based on simulations or propose selection methods using a predefined pilot estimator (see \cite{Croux2003}, \cite{WarwickJones2005}, \cite{GhoshBasu2016}, \cite{ghosh2019robust}). \cite{ribeiro2023} introduced an effective data-driven algorithm for selecting $\alpha$ for MDPDE and SMLE in beta regression models. \cite{Maluf2024} adapted this algorithm for LMDPDE and LSMLE. Here, we extend it to the proposed robust estimators for the zero-or-one inflated beta regression model. The algorithm is applied for estimating the parameters of the discrete and continuous parts separately. The algorithm selects the tuning constant as close to zero as possible (i.e., near MLE) while maintaining sufficient stability in the parameter estimates, thereby ensuring full efficiency when there is no contamination.

For completeness, we outline the steps of the algorithm presented by \cite{ribeiro2023}. Let $\alpha_0 = 0 < \alpha_1 < \alpha_2 < \cdots \leq 1$ be an ordered grid of values for $\alpha$, and let $z_{\alpha_k}$ be the vector of standardized estimates with tuning constant $\alpha_k$, i.e.,
	$$z_{\alpha_k}=\left(\frac{\widehat{\delta}^{1}_{\alpha_k}}{\sqrt{n}\ \mbox{se}(\widehat{\delta}^{1}_{\alpha_k})},\cdots, \frac{\widehat{\delta}^{p}_{\alpha_k}}{\sqrt{n}\ \mbox{se}(\widehat{\delta}^{p}_{\alpha_k})}\right)^\top,$$
	where $\mbox{se}(\cdot)$ denotes the asymptotic standard error.
	
	We define the standardized quadratic variation (SQV) for  $\widehat{{\boldsymbol\delta}}_{\alpha_k}$ as
	$\mbox{SQV}_{\alpha_k} = p^{-1}|| z_{\alpha_k} - z_{\alpha_{k+1}}||.$
	If $\mbox{SQV}_{\alpha_k}$ is small, the estimation with $\alpha=\alpha_k$ and $\alpha=\alpha_{k+1}$ are similar. 
	
	\begin{enumerate}
	
	\item Define an ordered, equally spaced grid for $\alpha$: $\alpha_0 = 0 < \alpha_1 < \alpha_2 < \cdots \leq \alpha_{m_1}$. 
	
	\item  if the stability condition $\mbox{SQV}_{\alpha_k} < L$, where $L>0$ is a pre-defined threshold, holds for all $k=0, 1, \ldots, m_1-1$, set the optimal value of $\alpha$ at $\alpha^{*}= \alpha_0 = 0$; otherwise, set $\alpha_{\rm start}$ at the next point in the grid after the smallest $\alpha_k$ for which the stability condition is not satisfied; 
	
	\item  define a new ordered, equally spaced grid for $\alpha$ starting from $\alpha_{\text{start}}$: $\alpha_0 = \alpha_{\text{start}} < \alpha_1 < \alpha_2 < \cdots < \alpha_{m}$, where $\alpha_m \leq \alpha_{\text{max}}$;
	
	\item  if the stability condition is satisfied for all $k=0, 1, \ldots, m-1$, set the optimal value of $\alpha$ at $\alpha^{*}= \alpha_0 = \alpha_{ \rm start}$;   otherwise, set $\alpha_{\rm start}$ at the next point in the grid after the smallest $\alpha_k$ for which the stability condition does not hold;
	
	\item repeat steps 3-4 until achieving stability of the estimates in the current grid or reaching the maximum value of $\alpha$, $\alpha_{\rm max}$;
	
	\item if $\alpha_{\text{max}}$ is reached without stability in the last grid, repeat steps 3-5 with $\alpha_{\rm start} = 0$; 
		
	\item if $\alpha_{\text{max}}$ is reached again without stability in the last grid,  set $\alpha^{*}= 0$.

\end{enumerate}

The algorithm checks if the stability condition is met for $\alpha \leq \alpha_{m_1}$ on a grid starting from $\alpha=0$. If satisfied, it chooses $\alpha^\ast = 0$ as the optimal value. If not, it seeks a sequence of $m$ consecutive $\alpha$ values that meet the stability condition. If unsuccessful, it restarts from $\alpha=0$. If the optimal $\alpha$ is not found, it returns $\alpha^\ast = 0$ (MLE).

The algorithm for choosing the tuning constant for the robust estimators of $\bm{\theta}$, i.e., the parameters of the continuous component, depend on the asymptotic standard errors used for computing SQV, which, in turn, depend on $\bm{\kappa}$. To make the choice of the tuning constant for the continuous part independent of the discrete part's parameter estimates, we suggest using standard errors from a beta regression model fitted to the observations in $(0,1)$.

Following \cite{ribeiro2023}, for the tuning constant of the continuous part, we suggest setting $\alpha_{\text{max}}=0.5$, $\alpha_{m_1}=0.2$, $L=0.02$, the grid spacings at $0.02$ (hence $m_1 = 10$), and $m=3$.  For the tuning constant of the discrete part, we suggest setting $\alpha_{\text{max}}=1$, $\alpha_{m_1}=0.5$, $L=0.02$, the grid spacings at $0.05$ (hence $m_1 = 11$), and $m=3$. These values are employed in our simulations and application.

We have implemented a user-friendly function for fitting M-LSE and M-LME in the R software \citep{R2024}, available in the GitHub repository at \url{https://github.com/ffqueiroz/robustinflatedbetareg}. The tuning constant is selected using the data-driven algorithm. The code also provides functions to obtain residuals and envelope plots for fitting the zero-inflated beta regression model. Currently, the residual function offers two types of residuals: the randomized quantile residual \citep{dunn1996} and a `by-part' residual, which separates the residuals for the discrete and continuous parts (deviance residual \citep{McCullaghNelder1989} for the discrete part and standardized weighted residual 2 \citep{Espinheira2008} for the continuous part). The envelope function utilizes the randomized quantile residual. These functions are employed in this paper for computing M-LSE and M-LME and the \texttt{gamlss} package is used for fitting inflated beta models via MLE.

It is important to emphasize that the proposed method does not simply split the data into continuous and discrete parts to apply existing robust techniques separately. While point estimates are computed separately for each component, the standard errors are derived jointly.  This allows for interval estimation and hypothesis testing to be performed within the unified and novel framework introduced in this work.

\section{Empirical results}\label{empiricalresults}

\begin{figure}[!h]
\centering
\includegraphics[scale=0.5]{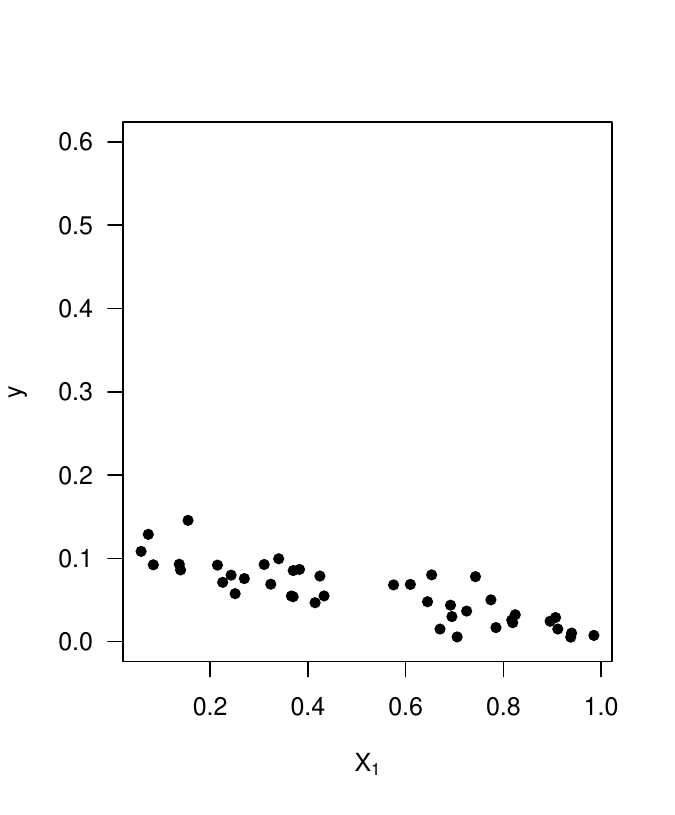}
\includegraphics[scale=0.5]{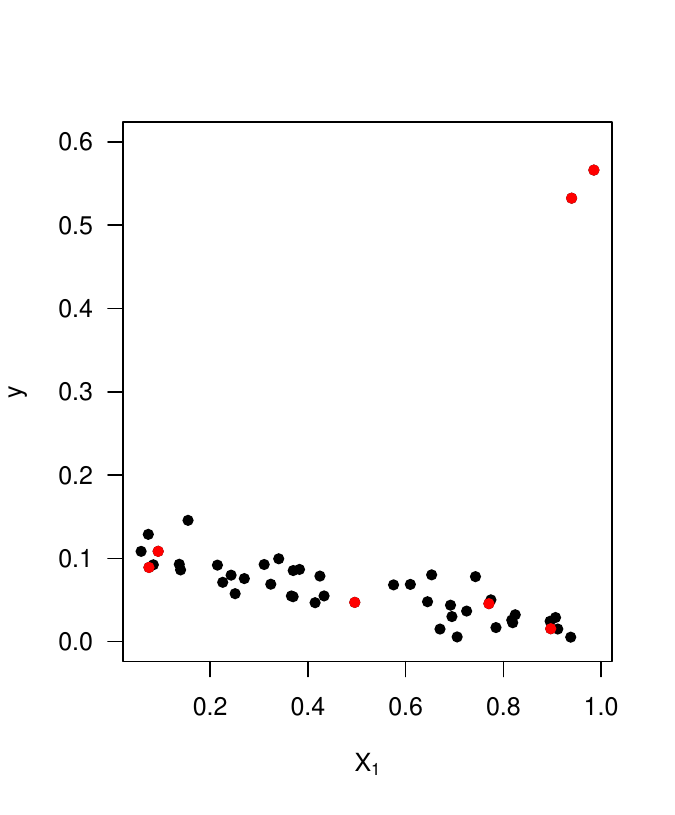}
\includegraphics[scale=0.5]{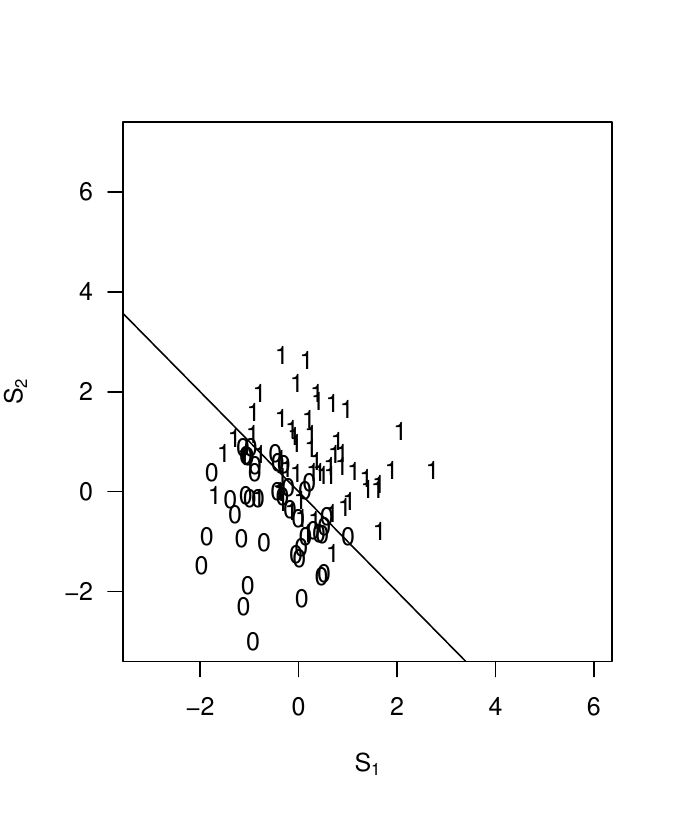}
\includegraphics[scale=0.5]{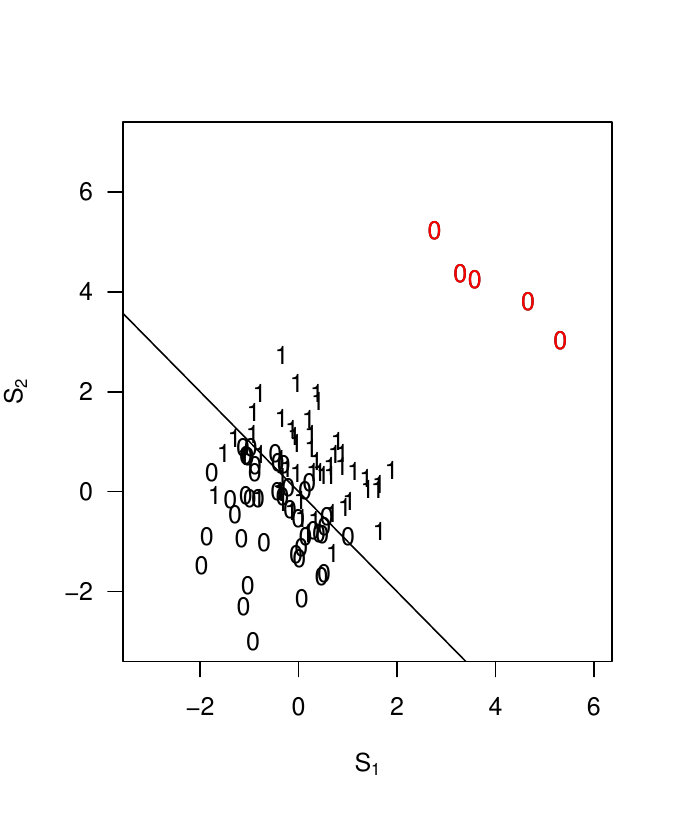}
\caption{Scatter plots of the non-contaminated sample (left) and the contaminated sample (right) for the continuous part (top row) and the discrete part (bottom row). The red points represent contaminated data.}\label{contamination}
\end{figure}

We generated $n=100$ observations of a zero-inflated beta regression model with 
\begin{align}\label{linearpred}
\begin{split}
 \log\left( \dfrac{\vartheta_i}{1-\vartheta_i} \right) = \kappa_1 + \kappa_2 \mathcal{S}_{i1} + \kappa_3 \mathcal{S}_{i2} ,\quad
\log\left( \dfrac{\mu_i}{1-\mu_i} \right)  = \beta_1 + \beta_2 \mathcal{X}_{i1}  , \quad
\log (\phi_i) = \gamma,
\end{split} 
\end{align}
where $\kappa_1 = 0$, $\kappa_2 = 2$, $\kappa_3 = 2$, $\beta_1 = -1.8$, $\beta_2 = -2$, and $\gamma=4.5$. The covariates $\mathcal{S}_{1}$ and $\mathcal{S}_{2}$ are obtained as independent realizations of a random variable $Z \sim \mathcal{N}(0, 1)$, and $\mathcal{X}_{1}$ is obtained from a random variable $U \sim \text{U}(0,1)$. These data will be identified as the non-contaminated data. We then contaminate the data as follows:
\begin{itemize}
\item[i.] (Continuous part) We contaminate $5\%$ of the observations in the continuous part; that is, observations such that $y \in (0,1)$. For example, in a sample of size $n=100$, suppose that $n^\dagger = 40$ observations are non-zero; in this case, $40 \times 0.05 = 2$ observations in the continuous part will be contaminated. The contamination follows the approach of \cite{ribeiro2023} and is performed by replacing the observations generated with the $5\%$ lowest response means by observations generated from a beta regression model with mean $\mu_i' = (1+\mu_i)/2$ (and precision $\phi$). See the top row of Figure \ref{contamination}; the two points in red with the highest values of $y_i$ are contaminated data.
\item[ii.] (Discrete part) We contaminate $5\%$ of the observations following the approach of \cite{Croux2003}. The contamination is introduced by replacing the observations with the top $5\%$ highest values of $\vartheta_i$ with observations on $(0,1)$, i.e., substituting $\ind_{\{0\}} (y_i)$ with zero for these observations. Note that replacing $\ind_{\{0\}} (y_i)$ with zero, in the zero-inflated beta regression model, means that the value of $y_i$, initially expected to be zero, is changed to a value within $(0,1)$. We generate $y_i$ based on a beta regression model with mean $\mu_i = g_\mu^{-1}(\bm{\mathcal{X}}_i^\top \bm{\beta})$ and precision $\phi_i = g_\phi^{-1}(\bm{\mathcal{Z}}_i^\top \bm{\gamma})$, that is, they are generated according to the assumed model (without contamination) and thus follow the trend presented by the other observations; see the top row of Figure \ref{contamination}. In addition, we increase the values of the covariates $\mathcal{S}_{i1}$ and $\mathcal{S}_{i2}$ so that the misclassified observations lie on a hyperplane parallel to the true discriminant hyperplane, with a distance of $1.5\sqrt{p_0} = 1.5\sqrt{3}$ between them; see the bottom row of plots in Figure \ref{contamination}. This type of contamination produces high-leverage outliers, which significantly disrupt estimation via maximum likelihood.
\end{itemize}

We then fit the zero-inflated beta regression model by using the MLE and the proposed robust estimators. The results are presented in Table \ref{resultsapp}.

\begin{table}[!ht]
\footnotesize
\centering
\caption{Estimates of the parameters, along with their respective standard errors, considering the MLE and the robust estimators for the non-contaminated and contaminated data.} \label{resultsapp}
\begin{tabular}{rrrrrrrrrrrrrrrrrr}
\hline
& \multicolumn{5}{c}{MLE}              & & \multicolumn{5}{c}{M-LSE} & & \multicolumn{5}{c}{M-LME} \\ 
\cline{2-6} \cline{8-12} \cline{14-18}
& \multicolumn{2}{c}{Non-cont.}  & &  \multicolumn{2}{c}{Cont.} & & \multicolumn{2}{c}{Non-cont.}  & &  \multicolumn{2}{c}{Cont.} & & \multicolumn{2}{c}{Non-cont.}  & &  \multicolumn{2}{c}{Cont.}\\ 
\cline{2-3}\cline{5-6}\cline{8-9} \cline{11-12} \cline{14-15}\cline{17-18}
Par.       & Est.    & s.e.  & &  Est.    & s.e. & & Est.    & s.e. & &  Est.    & s.e.  & & Est.    & s.e. & &  Est.   & s.e.\\ 
\hline
$\kappa_1$ & $0.41$  & $0.29$ & & $0.04$  & $0.21$ & & $0.41$  & $0.29$ & & $0.30$  & $0.28$ & & $0.41$  & $0.29$ & & $0.30$  & $0.28$ \\  
$\kappa_2$ & $1.74$  & $0.46$ & & $-0.04$ & $0.19$ & & $1.74$  & $0.46$ & & $1.47$  & $0.42$ & & $1.74$  & $0.46$ & & $1.47$  & $0.42$ \\  
$\kappa_3$ & $2.08$  & $0.47$ & & $0.35$  & $0.18$ & & $2.08$  & $0.47$ & & $1.75$  & $0.42$ & & $2.08$  & $0.47$ & & $1.75$  & $0.42$ \\  
$\beta_1$  & $-1.84$ & $0.09$ & & $-1.89$ & $0.22$ & & $-1.84$ & $0.09$ & & $-1.92$ & $0.08$ & & $-1.84$ & $0.09$ & & $-1.92$ & $0.08$ \\  
$\beta_2$  & $-2.08$ & $0.19$ & & $-0.94$ & $0.38$ & & $-2.08$ & $0.19$ & & $-1.88$ & $0.18$ & & $-2.08$ & $0.19$ & & $-1.88$ & $0.18$ \\  
$\gamma$ & $4.80$  & $0.19$ & & $2.68$  & $0.21$ & & $4.80$  & $0.19$ & & $4.93$  & $0.19$ & & $4.80$  & $0.19$ & & $4.92$  & $0.19$ \\ 
\hline
\end{tabular}
\end{table}

For the non-contaminated data, the parameter estimates obtained from the robust estimators coincide with the maximum likelihood estimates; in fact, the data-driven algorithm selected $\alpha = 0$ for all robust estimators. However, the contamination significantly affects the maximum likelihood estimation. For instance, the estimates of $\kappa_2$ and $\beta_2$ change from $1.74$ and $-2.08$ (non-contaminated data) to $0.35$ and $-0.94$ (contaminated data), respectively. On the other hand, the robust estimates (for both estimators) for the contaminated data are close to those for the non-contaminated data. For the contaminated data, the optimal values obtained for the tuning constant for MDPDE (discrete part), and LMDPDE and LSMLE (continuous part) are $0.3$, $0.1$ and $0.1$, respectively.

%
%
%

We next evaluate the performance of the proposed estimators in three scenarios with sample sizes $n=100$ and $n=200$, and $3000$ Monte Carlo replications. Tuning constants are selected via the algorithm in Section \ref{robustestimators}. In each scenario, data are generated from a zero-inflated beta regression model with linear predictors as defined in \eqref{linearpred}. We then contaminate $5\%$ of observations in the continuous part (Scenario 1), discrete part (Scenario 2), and both discrete and continuous parts (Scenario 3), following the contamination methods outlined above. All simulations were performed using the R software \citep{R2024}. The empirical bias and the
root mean squared error (RMSE) of the maximum likelihood estimates for MLE, M-LME, and M-LSE across non-contaminated and contaminated data in the three scenarios are presented in Table \ref{cen12} (Scenarios 1 and 2) and Table \ref{cen3} (Scenario 3). Inspection of these tables shows that, for non-contaminated data, MLE presents bias close to zero and small RMSE, both of which approaches zero as the sample size increases. In contrast, the contamination significantly affects MLE.

\setlength{\tabcolsep}{4pt} 

\begin{table}[!ht]
\centering
\footnotesize
\caption{Bias and root mean squared error (RMSE) for the MLE, M-LME, and M-LSE across non-contaminated and contaminated data in Scenarios 1 and 2.}\label{cen12}
\begin{tabular}{lrrr rrrr rrrr rrrr}
  \toprule
  & \multicolumn{4}{c}{MLE} & \multicolumn{4}{c}{M-LME} & \multicolumn{4}{c}{M-LSE} \\ 
  \cmidrule(lr){2-5} \cmidrule(lr){6-9} \cmidrule(lr){10-13}
  & \multicolumn{2}{c}{Non-cont} & \multicolumn{2}{c}{Cont} & \multicolumn{2}{c}{Non-cont} & \multicolumn{2}{c}{Cont} & \multicolumn{2}{c}{Non-cont} & \multicolumn{2}{c}{Cont} \\
  \cmidrule(lr){2-3} \cmidrule(lr){4-5} \cmidrule(lr){6-7} \cmidrule(lr){8-9} \cmidrule(lr){10-11} \cmidrule(lr){12-13}
  & bias & RMSE & bias & RMSE & bias & RMSE & bias & RMSE & bias & RMSE & bias & RMSE \\
  \midrule
  \multicolumn{13}{l}{\textit{Scenario 1}}\\
 \hline
    \multicolumn{13}{l}{$n=100$}\\
  \hline
  $\kappa_1$ & $0.01$ & $0.31$ & $0.01$ & $0.31$ & $0.01$ & $0.31$ & $0.01$ & $0.31$ & $0.01$ & $0.31$ & $0.01$ & $0.31$ \\
        $\kappa_2$ & $0.15$ & $0.54$ & $0.15$ & $0.54$ & $0.15$ & $0.54$ & $0.15$ & $0.54$ & $0.15$ & $0.54$ & $0.15$ & $0.54$ \\
        $\kappa_3$ & $0.16$ & $0.55$ & $0.16$ & $0.55$ & $0.16$ & $0.55$ & $0.16$ & $0.55$ & $0.16$ & $0.55$ & $0.16$ & $0.55$ \\
        $\beta_1$ & $-0.00$ & $0.10$ & $-0.04$ & $0.10$ & $-0.00$ & $0.10$ & $-0.00$ & $0.10$ & $-0.00$ & $0.10$ & $-0.00$ & $0.10$ \\
        $\beta_2$ & $0.00$ & $0.21$ & $1.11$ & $1.13$ & $0.00$ & $0.22$ & $0.00$ & $0.23$ & $0.00$ & $0.22$ & $0.01$ & $0.24$ \\
        $\gamma$ & $0.06$ & $0.22$ & $-1.86$ & $1.87$ & $0.06$ & $0.22$ & $0.04$ & $0.23$ & $0.06$ & $0.22$ & $0.04$ & $0.26$ \\
  \hline
  \multicolumn{13}{l}{$n=200$}\\
  \hline
$\kappa_1$ & $-0.00$ & $0.20$ & $-0.00$ & $0.20$ & $-0.00$ & $0.20$ & $-0.00$ & $0.20$ & $-0.00$ & $0.20$ & $-0.00$ & $0.20$ \\
        $\kappa_2$ & $0.07$ & $0.33$ & $0.07$ & $0.33$ & $0.07$ & $0.33$ & $0.07$ & $0.33$ & $0.07$ & $0.33$ & $0.07$ & $0.33$ \\
        $\kappa_3$ & $0.08$ & $0.35$ & $0.08$ & $0.35$ & $0.08$ & $0.35$ & $0.08$ & $0.35$ & $0.08$ & $0.35$ & $0.08$ & $0.35$ \\
        $\beta_1$ & $0.00$ & $0.08$ & $-0.09$ & $0.12$ & $0.00$ & $0.08$ & $0.00$ & $0.08$ & $0.00$ & $0.08$ & $-0.00$ & $0.08$ \\
        $\beta_2$ & $-0.00$ & $0.16$ & $1.19$ & $1.20$ & $-0.00$ & $0.16$ & $-0.00$ & $0.17$ & $-0.00$ & $0.16$ & $0.00$ & $0.17$ \\
        $\gamma$ & $0.03$ & $0.15$ & $-1.89$ & $1.89$ & $0.03$ & $0.15$ & $0.01$ & $0.16$ & $0.03$ & $0.15$ & $0.02$ & $0.16$ \\
  \midrule
  \multicolumn{13}{l}{\textit{Scenario 2}}\\
 \hline
    \multicolumn{13}{l}{$n=100$}\\
  \hline
$\kappa_1$ & $0.01$ & $0.31$ & $-0.28$ & $0.32$ & $0.01$ & $0.31$ & $-0.01$ & $0.30$ & $0.01$ & $0.31$ & $-0.01$ & $0.30$ \\
        $\kappa_2$ & $0.15$ & $0.54$ & $-1.85$ & $1.85$ & $0.15$ & $0.54$ & $-0.01$ & $0.56$ & $0.15$ & $0.54$ & $-0.01$ & $0.56$ \\
        $\kappa_3$ & $0.16$ & $0.55$ & $-1.70$ & $1.70$ & $0.16$ & $0.55$ & $-0.00$ & $0.56$ & $0.16$ & $0.55$ & $-0.00$ & $0.56$ \\
        $\beta_1$ & $-0.00$ & $0.10$ & $-0.00$ & $0.09$ & $-0.00$ & $0.10$ & $-0.00$ & $0.09$ & $-0.00$ & $0.10$ & $-0.00$ & $0.09$ \\
        $\beta_2$ & $0.00$ & $0.21$ & $0.00$ & $0.21$ & $0.00$ & $0.22$ & $0.00$ & $0.21$ & $0.00$ & $0.22$ & $0.00$ & $0.21$ \\
        $\gamma$ & $0.06$ & $0.22$ & $0.05$ & $0.21$ & $0.06$ & $0.22$ & $0.05$ & $0.21$ & $0.06$ & $0.22$ & $0.06$ & $0.21$ \\
    \hline
  \multicolumn{13}{l}{$n=200$}\\
  \hline
  $\kappa_1$ & $-0.00$ & $0.20$ & $-0.30$ & $0.32$ & $-0.00$ & $0.20$ & $-0.01$ & $0.20$ & $-0.00$ & $0.20$ & $-0.01$ & $0.20$ \\
        $\kappa_2$ & $0.07$ & $0.33$ & $-1.84$ & $1.85$ & $0.07$ & $0.33$ & $-0.07$ & $0.34$ & $0.07$ & $0.33$ & $-0.07$ & $0.34$ \\
        $\kappa_3$ & $0.08$ & $0.35$ & $-1.85$ & $1.85$ & $0.08$ & $0.35$ & $-0.07$ & $0.36$ & $0.08$ & $0.35$ & $-0.07$ & $0.36$ \\
        $\beta_1$ & $0.00$ & $0.08$ & $0.00$ & $0.07$ & $0.00$ & $0.08$ & $0.00$ & $0.07$ & $0.00$ & $0.08$ & $0.00$ & $0.07$ \\
        $\beta_2$ & $-0.00$ & $0.16$ & $-0.00$ & $0.15$ & $-0.00$ & $0.16$ & $-0.00$ & $0.15$ & $-0.00$ & $0.16$ & $-0.00$ & $0.15$ \\
        $\gamma$ & $0.03$ & $0.15$ & $0.03$ & $0.14$ & $0.03$ & $0.15$ & $0.03$ & $0.14$ & $0.03$ & $0.15$ & $0.03$ & $0.14$ \\
  \bottomrule
\end{tabular}
\end{table}

\begin{table}[!ht]
\centering
\footnotesize
\caption{Mean estimates (mean) and standard errors (se) for the MLE, M-LME, and M-LSE across non-contaminated and contaminated data in Scenarios 3.}\label{cen3}
\begin{tabular}{lrrr rrrr rrrr rrrr}
  \toprule
  & \multicolumn{4}{c}{MLE} & \multicolumn{4}{c}{M-LME} & \multicolumn{4}{c}{M-LSE} \\ 
  \cmidrule(lr){2-5} \cmidrule(lr){6-9} \cmidrule(lr){10-13}
  & \multicolumn{2}{c}{Non-cont} & \multicolumn{2}{c}{Cont} & \multicolumn{2}{c}{Non-cont} & \multicolumn{2}{c}{Cont} & \multicolumn{2}{c}{Non-cont} & \multicolumn{2}{c}{Cont} \\
  \cmidrule(lr){2-3} \cmidrule(lr){4-5} \cmidrule(lr){6-7} \cmidrule(lr){8-9} \cmidrule(lr){10-11} \cmidrule(lr){12-13}
  & bias & RMSE & bias & RMSE & bias & RMSE & bias & RMSE & bias & RMSE & bias & RMSE \\
  \midrule
    \multicolumn{13}{l}{$n=100$}\\
  \hline
$\kappa_1$ & $0.01$ & $0.31$ & $-0.28$ & $0.32$ & $0.01$ & $0.31$ & $-0.01$ & $0.30$ & $0.01$ & $0.31$ & $-0.01$ & $0.30$ \\
        $\kappa_2$ & $0.15$ & $0.54$ & $-1.85$ & $1.85$ & $0.15$ & $0.54$ & $-0.01$ & $0.56$ & $0.15$ & $0.54$ & $-0.01$ & $0.56$ \\
        $\kappa_3$ & $0.16$ & $0.55$ & $-1.70$ & $1.70$ & $0.16$ & $0.55$ & $-0.00$ & $0.56$ & $0.16$ & $0.55$ & $-0.00$ & $0.56$ \\
        $\beta_1$ & $-0.00$ & $0.10$ & $-0.04$ & $0.10$ & $-0.00$ & $0.10$ & $-0.00$ & $0.10$ & $-0.00$ & $0.10$ & $-0.00$ & $0.10$ \\
        $\beta_2$ & $0.00$ & $0.21$ & $1.05$ & $1.07$ & $0.00$ & $0.22$ & $0.00$ & $0.22$ & $0.00$ & $0.22$ & $0.01$ & $0.23$ \\
        $\gamma$ & $0.06$ & $0.22$ & $-1.79$ & $1.80$ & $0.06$ & $0.22$ & $0.03$ & $0.22$ & $0.06$ & $0.22$ & $0.04$ & $0.23$ \\
      \hline
  \multicolumn{13}{l}{$n=200$}\\
  \hline
$\kappa_1$ & $-0.00$ & $0.20$ & $-0.30$ & $0.32$ & $-0.00$ & $0.20$ & $-0.01$ & $0.20$ & $-0.00$ & $0.20$ & $-0.01$ & $0.20$ \\
        $\kappa_2$ & $0.07$ & $0.33$ & $-1.84$ & $1.85$ & $0.07$ & $0.33$ & $-0.07$ & $0.34$ & $0.07$ & $0.33$ & $-0.07$ & $0.34$ \\
        $\kappa_3$ & $0.08$ & $0.35$ & $-1.85$ & $1.85$ & $0.08$ & $0.35$ & $-0.07$ & $0.36$ & $0.08$ & $0.35$ & $-0.07$ & $0.36$ \\
        $\beta_1$ & $0.00$ & $0.08$ & $-0.07$ & $0.10$ & $0.00$ & $0.08$ & $0.00$ & $0.08$ & $0.00$ & $0.08$ & $-0.00$ & $0.08$ \\
        $\beta_2$ & $-0.00$ & $0.16$ & $1.07$ & $1.08$ & $-0.00$ & $0.16$ & $-0.00$ & $0.16$ & $-0.00$ & $0.16$ & $0.00$ & $0.16$ \\
        $\gamma$ & $0.03$ & $0.15$ & $-1.82$ & $1.82$ & $0.03$ & $0.15$ & $0.01$ & $0.15$ & $0.03$ & $0.15$ & $0.02$ & $0.15$ \\
  \bottomrule
\end{tabular}
\end{table}

In Scenario 1, the contamination is introduced only in observations that fall on $(0,1)$, impacting only the maximum likelihood estimates of the $\beta$'s and $\gamma$; see Table \ref{cen12}, Scenario 1. The MLEs of $\beta_2$ and $\gamma$ are notably influenced by the contamination, displaying severe bias. For example, with $n=100$, the MLE for $\beta_2$ under non-contaminated data shows negligible bias (zero up to two decimal places), whereas under contamination, the bias increases to $1.11$. 
In contrast, the robust estimators perform well on both contaminated and non-contaminated data. For non-contaminated data, their performance closely matches that of MLE, as the selection algorithm for the tuning constant selects $\alpha=0$ (equivalent to MLE) when no outliers are detected. In contaminated data, these estimators remain unaffected by outliers, yielding results similar to MLE in non-contaminated data. The M-LME and M-LSE for $\beta_1$, $\beta_2$, and $\gamma$ exhibit near-zero bias and low RMSE, outperforming MLE under contamination. For example, with $n=200$, the bias and RMSE of M-LME for $\gamma$ are $0.01$ and $0.16$, respectively, compared to $-1.89$ and $1.89$ for MLE. As expected, the robust estimates of the $\kappa$'s are similar for both contaminated and non-contaminated data and align with MLE, indicating that the $\alpha$ selection algorithm correctly chose $\alpha=0$ for the discrete part.

In Scenario 2, contamination occurs only in the discrete part, affecting only the MLE for the $\kappa$'s; see Table \ref{cen12}, Scenario 2. For the contaminated data, the maximum likelihood estimates for the $\kappa$'s deviate significantly from its true value. For instance, the bias of the MLE for $\kappa_3$ under non-contaminated data is $0.16$ ($n=200$), whereas under contamination, the bias shifts to $-1.70$. Note that, under contaminated data, the biases of $\kappa_2$ and $\kappa_3$ are negative and, in absolute value, they are close to their true parameter values. Additionally, the variability of these estimates is small, as the RMSE is nearly equal to the square root of the squared bias. Hence, the estimates of $\kappa_2$ and $\kappa_3$ are near zero. This behavior is expected, as \cite{Croux2002} demonstrated that MLEs for the $\kappa$'s (except for the intercept) tend to zero in the presence of contamination. The robust estimators provide reasonable estimates under both contaminated and non-contaminated data. Without contamination, the selection algorithm typically chooses $\alpha=0$. Under contamination, the robust estimators behave similarly to MLE under non-contaminated data, showing that outliers do not influence the proposed robust estimations.

In Scenario 3, contamination occurs in both the continuous and discrete parts. The performance of the robust estimators is similar to that observed in Scenarios 1 and 2: in the absence of contamination, the tuning parameter selection algorithm chooses $\alpha=0$ in most cases, and the estimates closely align with the maximum likelihood estimates; under contamination, the robust estimators behave like the MLEs under non-contaminated data, yielding estimates with bias close to zero and small RMSE; see Table \ref{cen3}.

The algorithm for selecting the tuning parameter for the robust estimators performed exceptionally well. Table \ref{alphavalues} shows the mean and standard deviation (sd) of the optimal $\alpha$ values selected by the data-driven algorithm for both the continuous and discrete parts of the M-LME and M-LSE across all the three scenarios. 

\begin{table}[!ht]
\centering
\footnotesize
\caption{Mean and standard deviation (sd) of the optimal $\alpha$ values selected by the data-driven algorithm for both the continuous and discrete parts of M-LME and M-LSE across Scenarios 1, 2, and 3.}\label{alphavalues}
\begin{tabular}{lrrr rrrr rrrr rrrr}
  \toprule
  & \multicolumn{4}{c}{Scenario 1} & \multicolumn{4}{c}{Scenario 2} & \multicolumn{4}{c}{Scenario 3} \\ 
  \cmidrule(lr){2-5} \cmidrule(lr){6-9} \cmidrule(lr){10-13}
  & \multicolumn{2}{c}{Non-cont} & \multicolumn{2}{c}{Cont} & \multicolumn{2}{c}{Non-cont} & \multicolumn{2}{c}{Cont} & \multicolumn{2}{c}{Non-cont} & \multicolumn{2}{c}{Cont} \\
  \cmidrule(rr){2-3} \cmidrule(rr){4-5} \cmidrule(rr){6-7} \cmidrule(rr){8-9} \cmidrule(rr){10-11} \cmidrule(rr){12-13}
  & mean & sd & mean & sd & mean & sd & mean & sd & mean & sd & mean & sd \\
  \midrule
    \multicolumn{13}{l}{$n=100$}\\
  \hline
discrete part & $0.00$ & $0.00$ & $0.00$ & $0.00$ & $0.00$ & $0.00$ & $0.29$ & $0.04$ & $0.00$ & $0.00$ & $0.29$ & $0.04$ \\ 
continuous part (LMDPDE) & $0.00$ & $0.01$ & $0.12$ & $0.02$ & $0.00$ & $0.01$ & $0.00$ & $0.00$ & $0.00$ & $0.01$ & $0.12$ & $0.02$ \\ 
continuous part (LSMLE) & $0.00$ & $0.01$ & $0.11$ & $0.02$ & $0.00$ & $0.01$ & $0.00$ & $0.01$ & $0.00$ & $0.01$ & $0.11$ & $0.02$ \\ 
      \hline
  \multicolumn{13}{l}{$n=200$}\\
  \hline
discrete part & $0.00$ & $0.00$ & $0.00$ & $0.00$ & $0.00$ & $0.00$ & $0.28$ & $0.03$ & $0.00$ & $0.00$ & $0.28$ & $0.03$ \\ 
continuous part (LMDPDE) & $0.00$ & $0.00$ & $0.12$ & $0.01$ & $0.00$ & $0.00$ & $0.00$ & $0.00$ & $0.00$ & $0.00$ & $0.12$ & $0.01$ \\ 
continuous part (LSMLE) & $0.00$ & $0.00$ & $0.11$ & $0.01$ & $0.00$ & $0.00$ & $0.00$ & $0.00$ & $0.00$ & $0.00$ & $0.11$ & $0.01$ \\ 
  \bottomrule
\end{tabular}
\end{table}

For non-contaminated data, the algorithm selects $\alpha=0$ for nearly all samples when $n=100$ and for all samples when $n=200$ (the mean and standard deviation of the optimal $\alpha$ values are both zero for $n=200$). For contaminated data, when contamination occurs in the discrete part, the algorithm typically selects $\alpha=0$ for the robust estimators of the continuous part, and vice-versa. It tends to choose $\alpha$ around $0.28$ for the robust estimators of the discrete part and around $0.11$ for the continuous part. Thus, in all three scenarios, the algorithm successfully identifies the need for a robust method when contamination is present and selects MLE when no contamination exists. This ability is crucial for achieving an optimal balance between efficiency and robustness, ensuring maximum efficiency in the absence of outliers. 

We assess the efficiency of the proposed estimators with and without contamination using total mean squared errors (TMSE). Table \ref{EQMTSIM1} presents the TMSE ratios for MLE, M-LME, and M-LSE under Scenarios 1, 2, and 3. Without contamination, all TMSE ratios equal one. For the contaminated data, the robust estimators show significantly lower TMSEs than MLE, especially for larger samples ($n=200$). For instance, in Scenario 3 with $n=200$, the TMSE of MLE is about 34 times larger than that of M-LME and M-LSE. The M-LME and M-LSE behave similarly, with TMSE ratios either equal to or very close to one across all scenarios.

\begin{table}[!ht]
\centering
\footnotesize
\def\arraystretch{1.3}
\caption{Ratio of the total mean squared errors of the estimators under Scenarios 1, 2, and 3.}\label{EQMTSIM1}
\begin{tabular}{rrrrrrrrrrrrrrrrrr}
\hline
& \multicolumn{5}{c}{\shortstack{Scenario 1}} && \multicolumn{5}{c}{\shortstack{Scenario 2}} && \multicolumn{5}{c}{\shortstack{Scenario 3}} \\
\cline{2-6}\cline{8-12}\cline{14-18}
                                     & \multicolumn{2}{c}{\shortstack{Non-cont.}} & & \multicolumn{2}{c}{\shortstack{Cont.}} & & 
  \multicolumn{2}{c}{\shortstack{Non-cont.}} & &  \multicolumn{2}{c}{\shortstack{Cont.}} & & 
  \multicolumn{2}{c}{\shortstack{Non-cont.}} & &  \multicolumn{2}{c}{\shortstack{Cont.}} \\ 
\cline{2-3}\cline{5-6}\cline{8-9}\cline{11-12}\cline{14-15}\cline{17-18}
         $n$                           & $100$  & $200$   & &  $100$  & $200$    & &  $100$  & $200$   & &  $100$  & $200$    & &  $100$  & $200$   & &  $100$   & $200$ \\ 
  \midrule
$\frac{\text{MLE}}{\text{M-LSE}}   $ & $1.00$ & $1.00$  & &  $6.63$ & $15.90$  & &  $1.00$ & $1.00$  & &  $7.93$ & $20.93$  & &  $1.00$ & $1.00$  & &  $12.88$ & $33.75$ \\ 
$\frac{\text{MLE}}{\text{M-LME}}  $ & $1.00$ & $1.00$  & &  $6.77$ & $15.89$  & &  $1.00$ & $1.00$  & &  $7.94$ & $20.93$  & &  $1.00$ & $1.00$  & &  $12.98$ & $33.72$ \\ 
$\frac{\text{M-LSE}}{\text{M-LME}}$ & $1.00$ & $1.00$  & &  $1.02$ & $1.00$   & &  $1.00$ & $1.00$  & &  $1.00$ & $1.00$   & &  $1.00$ & $1.00$  & &  $1.01$  & $1.00$  \\ 
\hline
\end{tabular}
\end{table}

Finally, we evaluate the performance of the robust Wald-type test, proposed in Section \ref{robustestimators}, of $\mathcal{H}_0: \upsilon_j = \upsilon_j^{(0)}$ against the alternative $\mathcal{H}_1: \upsilon_j \neq \upsilon_j^{(0)}$. The null hypothesis $\mathcal{H}_0$ is rejected at the nominal level $\zeta$ when the statistic $W_j$ exceeds the $(1-\zeta)$ quantile of the $\chi^2_{1}$ distribution. We set $\zeta = 5\%$ and compute the empirical rejection rates of the robust Wald-type test under $\mathcal{H}_0$ based on MLE (standard Wald test) and the proposed robust estimators. This procedure was performed for Scenarios 1, 2, and 3, with the chosen values for $\upsilon_j^{(0)}$ fixed at the true parameter values in \eqref{linearpred}. The results are presented in Table \ref{testehipotesetab}.

\begin{table}[!ht]
\centering
\footnotesize
\caption{Empirical levels under $\mathcal{H}_0$ of the robust Wald-type test under Scenarios 1, 2, and 3 at the nominal level of $5\%$.} \label{testehipotesetab}
\begin{tabular}{rrrrrrrrrrrrrr}
\hline
     &           \multicolumn{6}{c}{\shortstack{Non-contaminated data}}             & &            \multicolumn{6}{c}{\shortstack{Contaminated data}}          \\ 
\cline{2-7}\cline{9-14}
    & $\kappa_1$ & $\kappa_2$ & $\kappa_3$& $\beta_1$ & $\beta_2$  & $\gamma$ & & $\kappa_1$ & $\kappa_2$ & $\kappa_3$& $\beta_1$ & $\beta_2$ & $\gamma$ \\ 
\hline
    \multicolumn{14}{l}{\textit{Scenario 1}}\\
\hline
    \multicolumn{14}{l}{$n=100$}\\
\hline
MLE          & $0.04$ & $0.04$ & $0.04$ & $0.06$ & $0.06$ & $0.07$       & & $0.04$ & $0.04$ & $0.04$ & $0.00$ & $0.95$ & $1.00$       \\ 
M-LSE        & $0.04$ & $0.04$ & $0.04$ & $0.06$ & $0.06$ & $0.07$       & & $0.04$ & $0.04$ & $0.04$ & $0.06$ & $0.08$ & $0.08$       \\ 
M-LME        & $0.04$ & $0.04$ & $0.04$ & $0.06$ & $0.06$ & $0.07$       & & $0.04$ & $0.04$ & $0.04$ & $0.06$ & $0.07$ & $0.08$       \\ 
\hline
    \multicolumn{14}{l}{$n=200$}\\
\hline
MLE          & $0.05$ & $0.04$ & $0.04$ & $0.06$ & $0.05$ & $0.06$       & & $0.05$ & $0.04$ & $0.04$ & $0.00$ & $1.00$ & $1.00$       \\ 
M-LSE        & $0.05$ & $0.04$ & $0.04$ & $0.06$ & $0.05$ & $0.06$       & & $0.05$ & $0.04$ & $0.04$ & $0.06$ & $0.07$ & $0.07$       \\ 
M-LME        & $0.05$ & $0.04$ & $0.04$ & $0.06$ & $0.05$ & $0.06$       & & $0.05$ & $0.04$ & $0.04$ & $0.06$ & $0.07$ & $0.06$       \\ 
\hline
    \multicolumn{14}{l}{\textit{Scenario 2}}\\
\hline
    \multicolumn{14}{l}{$n=100$}\\
\hline
MLE          & $0.04$ & $0.04$ & $0.04$ & $0.06$ & $0.06$ & $0.07$      & & $0.20$ & $1.00$ & $1.00$ & $0.06$ & $0.05$ & $0.07$       \\ 
M-LSE        & $0.04$ & $0.04$ & $0.04$ & $0.06$ & $0.06$ & $0.07$      & & $0.04$ & $0.09$ & $0.08$ & $0.06$ & $0.05$ & $0.06$       \\ 
M-LME        & $0.04$ & $0.04$ & $0.04$ & $0.06$ & $0.06$ & $0.07$      & & $0.04$ & $0.09$ & $0.08$ & $0.06$ & $0.05$ & $0.06$       \\ 
\hline
    \multicolumn{14}{l}{$n=200$}\\
\hline
MLE          & $0.05$ & $0.04$ & $0.04$ & $0.06$ & $0.05$ & $0.06$       & & $0.54$ & $1.00$ & $1.00$ & $0.06$ & $0.05$ & $0.06$       \\ 
M-LSE        & $0.05$ & $0.04$ & $0.04$ & $0.06$ & $0.05$ & $0.06$       & & $0.04$ & $0.09$ & $0.10$ & $0.05$ & $0.04$ & $0.05$       \\ 
M-LME        & $0.05$ & $0.04$ & $0.04$ & $0.06$ & $0.05$ & $0.06$       & & $0.04$ & $0.09$ & $0.10$ & $0.05$ & $0.04$ & $0.05$       \\ 
\hline
    \multicolumn{14}{l}{\textit{Scenario 3}}\\
\hline
    \multicolumn{14}{l}{$n=100$}\\
\hline
MLE          & $0.04$ & $0.04$ & $0.04$ & $0.06$ & $0.06$ & $0.07$       & & $0.20$ & $1.00$ & $1.00$ & $0.00$ & $0.93$ & $1.00$       \\ 
M-LSE        & $0.04$ & $0.04$ & $0.04$ & $0.06$ & $0.06$ & $0.07$       & & $0.04$ & $0.09$ & $0.08$ & $0.06$ & $0.07$ & $0.07$       \\ 
M-LME        & $0.04$ & $0.04$ & $0.04$ & $0.06$ & $0.06$ & $0.07$       & & $0.04$ & $0.09$ & $0.08$ & $0.06$ & $0.06$ & $0.07$       \\ 
\hline
    \multicolumn{14}{l}{$n=200$}\\
\hline
MLE          & $0.05$ & $0.04$ & $0.04$ & $0.06$ & $0.05$ & $0.06$       & & $0.54$ & $1.00$ & $1.00$ & $0.00$ & $1.00$ & $1.00$        \\ 
M-LSE        & $0.05$ & $0.04$ & $0.04$ & $0.06$ & $0.05$ & $0.06$       & & $0.04$ & $0.09$ & $0.10$ & $0.06$ & $0.06$ & $0.06$        \\ 
M-LME        & $0.05$ & $0.04$ & $0.04$ & $0.06$ & $0.05$ & $0.06$       & & $0.04$ & $0.09$ & $0.10$ & $0.05$ & $0.06$ & $0.06$        \\ 
\hline
\end{tabular}
\end{table}

Both the usual and robust Wald-type tests exhibit similar behavior when there is no contamination in the data. The empirical levels tend to be closer to the nominal level when the sample size is moderate/large ($n=200$). On the other hand, the performance of the usual Wald test is significantly compromised in the presence of contaminated data. In most cases, we observe type I error probabilities equal to one or zero; see the usual Wald tests for $\beta_1$ and $\kappa_2$ in Scenarios 1 and 2, respectively, for $n=100$ and contaminated data. This indicates that, in the presence of outliers, the usual Wald test should be avoided. As expected, this poor performance is observed exclusively in tests associated with the parameters of the contaminated part of the data, whether discrete or continuous. For example, in Scenario 1, in the presence of contamination, the usual Wald test performs well for parameters associated with the discrete part ($\kappa$'s), unlike for parameters associated with the continuous part ($\beta$'s and $\gamma$). In this scenario, the empirical levels for $\kappa_1$, $\kappa_2$, and $\kappa_3$, for $n=100$, are $4\%$, while those for $\beta_1$, $\beta_2$, and $\gamma$ are $0\%$, $95\%$, and $100\%$, respectively. Similarly, in Scenario 2, only the usual Wald tests associated with parameters from the discrete part are compromised, and in Scenario 3, the usual Wald tests for all parameters are compromised, as the contamination is applied to both the discrete and continuous parts.

In contrast, the robust Wald-type tests perform reasonably well both in the absence and presence of contamination. In most cases, when there is contamination, the empirical levels of the robust tests are close to the nominal levels. For example, in Scenario 3, for contaminated data, the tests for $\beta_1$, $\beta_2$, and $\gamma$, considering M-LME and $n=200$, show empirical levels of $5\%$, $6\%$, and $6\%$ respectively. In general, both robust estimators perform similarly. Additionally, the presence of contamination in the data slightly  distorts the type I error probabilities of the robust Wald-type tests beyond the expected levels. Thus, although these tests are much more appropriate than the usual Wald test, caution is needed when making decisions.

\section{Application} \label{application}

We now present a real data application concerning the COVID-19 case fatality rate (CFR) in $200$ cities of the state of Minas Gerais, Brazil. The response variable, denoted by \texttt{CFR}, is defined as the ratio of confirmed COVID-19 deaths to confirmed cases, accumulated up to October $1$, $2020$. The covariates are: \texttt{Pop}, the estimated population of the city in $2020$, and \texttt{HDI}, the Human Development Index of the city\footnote{The data were obtained from databases available at \url{https://www.seade.gov.br/}, \url{https://www.ibge.gov.br/}, and \url{https://www.undp.org/pt/brazil}.}. One city has a $100\%$ CFR, which is adjusted to $0.999$ for analysis purposes. It corresponds to observation $\#39$ and recorded a single COVID-19 case, which resulted in death. Additionally, $29.5\%$ of the cities reported a CFR of zero. Thus, we consider a zero inflated beta regression model with 
\begin{align*}
\begin{split}
 \log\left( \dfrac{\vartheta_i}{1-\vartheta_i} \right) &= \kappa_1 + \kappa_2 \times \texttt{Pop}_i + \kappa_3 \times \texttt{HDI}_i ,\\ 
\log\left( \dfrac{\mu_i}{1-\mu_i} \right)  &= \beta_1 + \beta_2 \times\texttt{Pop}_i + \beta_3 \times\texttt{HDI}_i  , \\ 
\log (\phi_i) &= \gamma_1 + \gamma_2 \times \texttt{Pop}_i + \gamma_3 \times\texttt{HDI}_i,
\end{split} 
\end{align*} 
for $i=1, \ldots, 200$. We fit the model using MLE and the two proposed robust estimators, M-LSE and M-LME. Since the estimates and standard errors from M-LSE and M-LME are similar, we report results only for M-LSE. Table \ref{estimates} provides the estimates, asymptotic standard errors, $z$-statistics (ratios of estimates to their asymptotic standard errors), and asymptotic $p$-values for Wald-type tests of nullity of coefficients. The table also shows results excluding observation $\#39$, the most prominent outlier. 

\renewcommand{\arraystretch}{0.9} 

\begin{table}[!ht]
\centering
\small
\caption{Estimates, asymptotic standard errors (se), $z$-stat, and asymptotic p-values for the full and reduced data for the MLE and the M-LSE.}\label{estimates}
\begin{tabular}{rrrrrrrrrr}
\hline
& \multicolumn{4}{c}{MLE for the full data}    & & \multicolumn{4}{c}{M-LSE for the full data}\\
\cline{2-5} \cline{7-10}
          & Estimate  & se & $z$-stat & $p$-value & & Estimate & se & $z$-stat & $p$-value \\ 
\hline
$\log[\vartheta/(1-\vartheta)]$ &&&&&&&&&\\
\texttt{Intercept} & $16.538$  & $3.712$ & $4.455$  & $0.000$  & & $16.538$ & $3.712$ & $4.455$  & $0.000$ \\ 
\texttt{Pop}        & $-1.564$  & $0.287$ & $-5.452$ & $0.000$  & & $-1.564$ & $0.287$ & $-5.452$ & $0.000$ \\ 
\texttt{HDI}       & $-5.194$  & $4.262$ & $-1.219$ & $0.225$  & & $-5.194$ & $4.262$ & $-1.219$ & $0.223$ \\ \\
$\log[\mu/(1-\mu)]$ &&&&&&&&&\\
\texttt{Intercept} & $-0.324$  & $0.971$ & $-0.333$ & $0.739$  & & $0.620$  & $0.809$ & $0.767$  & $0.443$ \\ 
\texttt{Pop}        & $-0.267$  & $0.096$ & $-2.774$ & $0.006$  & & $-0.130$ & $0.059$ & $-2.214$ & $0.027$ \\ 
\texttt{HDI}       & $-0.179$  & $2.045$ & $-0.087$ & $0.930$  & & $-3.882$ & $1.421$ & $-2.732$ & $0.006$ \\ \\
$\log \phi$ &&&&&&&&&\\
\texttt{Intercept} & $-0.515$  & $2.496$ & $-0.206$ & $0.837$  & & $-5.159$ & $1.821$ & $-2.834$ & $0.005$ \\ 
\texttt{Pop}        & $1.170$   & $0.099$ & $11.878$ & $0.000$  & & $0.668$  & $0.131$ & $5.100$  & $0.000$ \\ 
\texttt{HDI}      & $-11.102$ & $3.473$ & $-3.197$ & $0.002$  & & $3.954$  & $2.937$ & $1.346$  & $0.178$ \\ 
\hline
& \multicolumn{4}{c}{MLE without observation $\#39$}    & & \multicolumn{4}{c}{M-LSE without observation $\#39$}\\
\cline{2-5} \cline{7-10}
          & Estimate  & se & $z$-stat & $p$-value & & Estimate & se & $z$-stat & $p$-value \\ 
\hline
$\log[\vartheta/(1-\vartheta)]$ &&&&&&&&&\\
\texttt{Intercept} & $16.564$  & $3.717$ & $4.456$  & $0.000$  & & $16.564$ & $3.717$ & $4.456$  & $0.000$ \\ 
\texttt{Pop}      & $-1.594$  & $0.290$ & $-5.492$ & $0.000$  & & $-1.594$ & $0.290$ & $-5.493$ & $0.000$ \\ 
\texttt{HDI}     & $-4.800$  & $4.274$ & $-1.123$ & $0.263$  & & $-4.800$ & $4.274$ & $-1.123$ & $0.261$ \\ \\
$\log[\mu/(1-\mu)]$ &&&&&&&&&\\
\texttt{Intercept} & $0.706$   & $0.746$ & $0.946$  & $0.345$  & & $0.701$  & $0.823$ & $0.852$  & $0.394$ \\ 
\texttt{Pop}        & $-0.155$  & $0.055$ & $-2.810$ & $0.005$  & & $-0.154$ & $0.060$ & $-2.557$ & $0.011$ \\ 
\texttt{HDI}     & $-3.634$  & $1.337$ & $-2.719$ & $0.007$  & & $-3.641$ & $1.445$ & $-2.520$ & $0.012$ \\ \\
$\log \phi$ &&&&&&&&&\\
\texttt{Intercept} & $-4.950$  & $1.641$ & $-3.015$ & $0.003$  & & $-4.955$ & $1.822$ & $-2.719$ & $0.007$ \\ 
\texttt{Pop}      & $0.657$   & $0.126$ & $5.198$  & $0.000$  & & $0.657$  & $0.132$ & $4.969$  & $0.000$ \\ 
\texttt{HDI}     & $3.748$   & $2.747$ & $1.364$  & $0.174$  & & $3.759$  & $2.949$ & $1.275$  & $0.202$ \\ 
\hline
\end{tabular}
\end{table}

For the full data, the data-driven algorithm selects $\alpha = 0$ for the discrete component in both M-LSE and M-LME. Consequently, the estimates and asymptotic standard errors of the parameters of the zero-probability submodel based on these estimators are identical to those obtained using MLE. For this submodel, only the covariate \texttt{Pop} is statistically significant, with an estimate of $-1.567$. This suggests that population size is negatively associated with the probability of a zero CFR, which is intuitive--larger populations are more likely to experience deaths from COVID-19. For the continuous component, the algorithm  selects $\alpha=0.04$ for both M-LSE and M-LME, indicating the need for a robust fit. Indeed, the estimates and asymptotic standard errors (and, consequently, the $p$-values) for the parameters of the conditional mean and precision submodels by M-LSE differ notably from those obtained using MLE. In the conditional mean submodel, \texttt{HDI} is not statistically significant when MLE is employed but is found to be significant when using M-LSE, with the estimate shifting from $-0.179$ to $-3.882$. For the conditional precision submodel, notable differences are also observed. The MLE indicates that \texttt{HDI} is statistically significant; in contrast, it does not show a significant effect when using M-LSE. 

\begin{figure}[!h]
\centering
\includegraphics[scale=0.5]{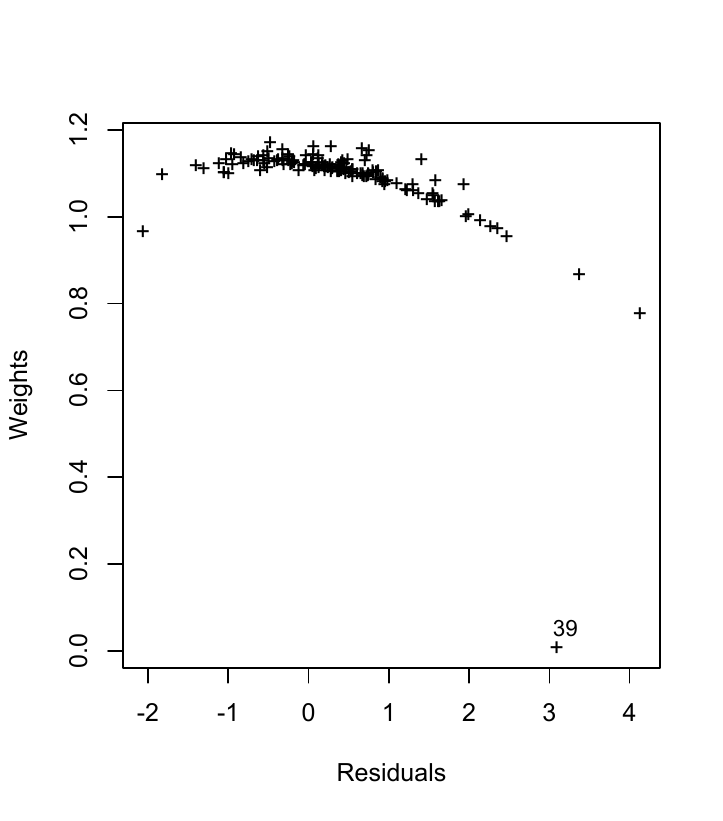}
\caption{Scatter plot of weights versus quantile residuals for M-LSE (continuous part).}\label{weights}
\end{figure}

The exclusion of observation $\#39$ has only a slight impact on M-LSE, but highly influences MLE-based inference. For example, when case $\#39$ is excluded, the \texttt{HDI} coefficient estimate shifts from $-0.179$ to $-3.634$ under MLE, and from $-3.882$ to $-3.641$ under M-LSE. The inferences based on M-LSE for both the full data and the data without the outlier are similar and closely align with that based on MLE for the data excluding observation $\#39$. In fact, case $\#39$ receives small weight in the M-LSE estimation process; see Figure \ref{weights}.

\begin{figure}[!h]
\centering
\includegraphics[scale=0.5]{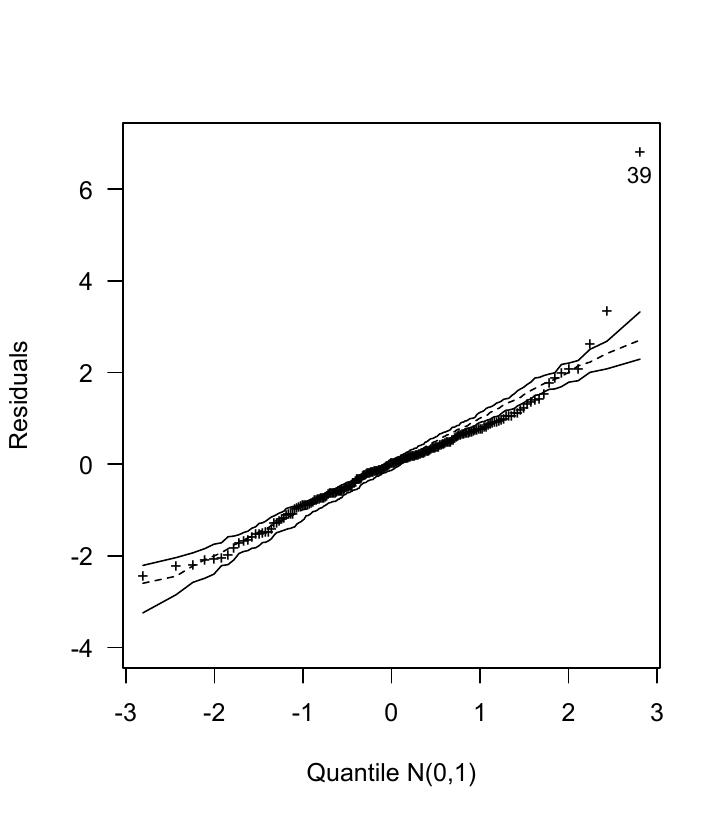}
\includegraphics[scale=0.5]{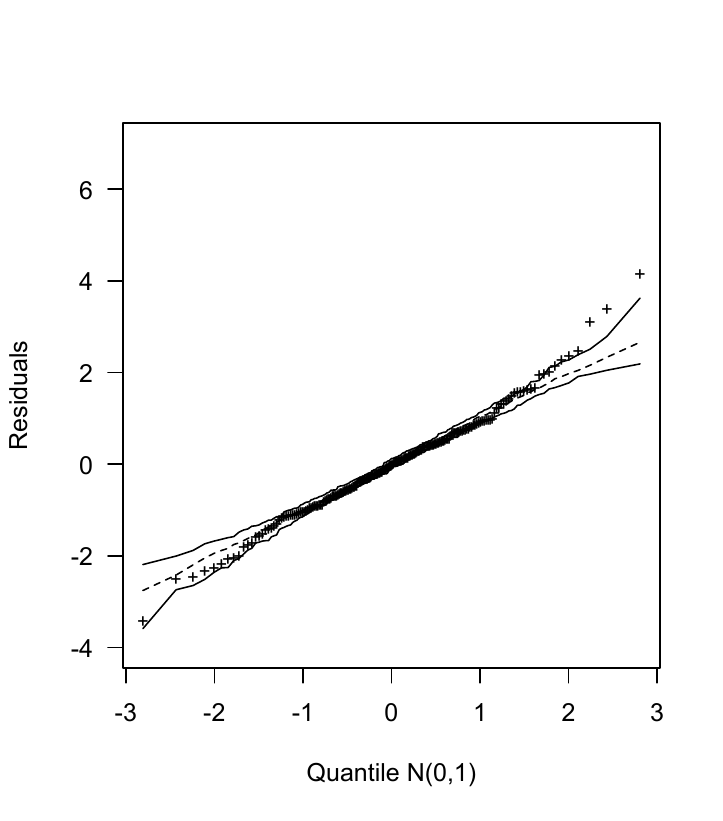}
\includegraphics[scale=0.5]{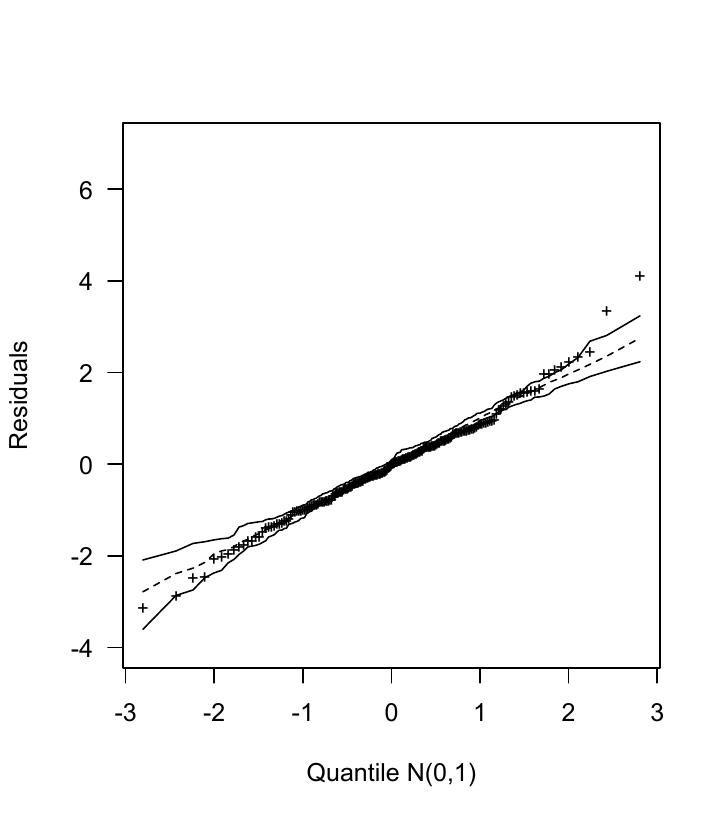}
\includegraphics[scale=0.5]{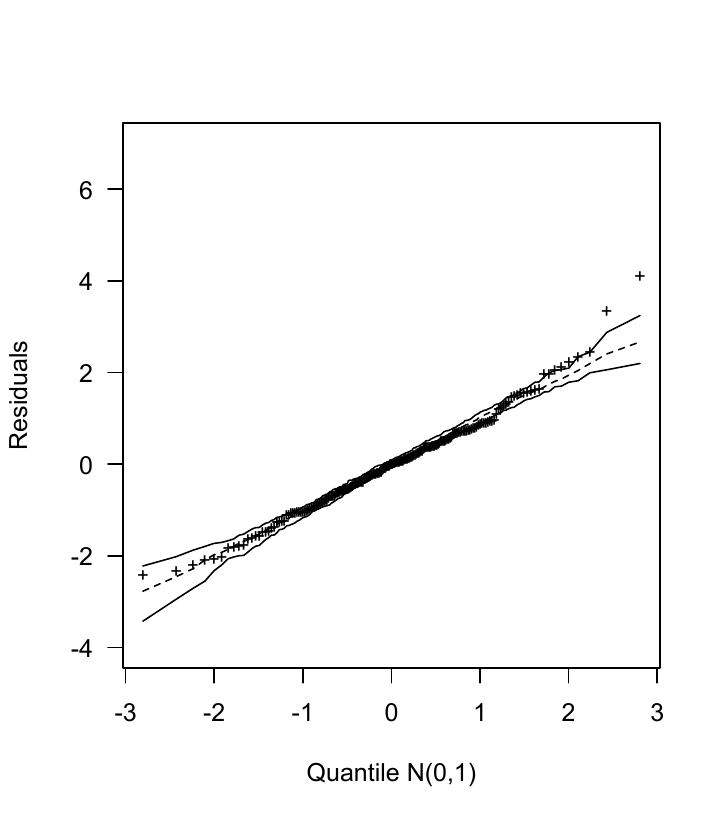}
\caption{Plots of the quantile residuals with simulated envelopes for the fits based on MLE (left) and M-LSE (right) for the full data (first row) and the data without observation $\#39$ (second row).}\label{res}
\end{figure}

Figure \ref{res} presents the plots of the quantile residuals with simulated envelopes for the fits based on MLE and M-LSE for the full data and the data without case $\#39$. The plot for the fit of the MLE for the full data clearly evidences a lack of fit and highlights observation $\#39$ as an outlier. In contrast, the other three plots indicate reasonable fits.

Overall, this empirical application illustrates that a single observation can have a significant impact on maximum likelihood-based inference, potentially producing misleading conclusions. In contrast, robust estimators reduce the influence of outliers, yielding more reliable results.

\section{Concluding remarks}

Although there are different distributions to model continuous proportions, the beta distribution is by far the most employed. When the data contains observations at the boundaries of the unit interval, inflated beta regression models are widely used. It is well-known that the maximum likelihood estimation is not robust against outliers. When outliers are present in the data, one may change the model for a more flexible one or change the inference method. Flexible models usually have more parameters than the inflated beta model, which sometimes do not have clear interpretations, being more complex and requiring more computational efforts; see, for instance, \cite{QueirozFerrari2010}. Our approach is to keep the inflated beta models, while changing the estimation process by a robust one. 

We proposed robust estimators for zero-or-one inflated beta regression models. The new estimators overcome the lack of robustness of the maximum likelihood estimator while keeping good asymptotic properties. We also proposed an algorithm that determines an appropriate tuning constant based on the robustness required by the data. Only few studies address robust inference in inflated models and, to the best of our knowledge, ours is the first to consider robust inference methods for inflated models for bounded continuous data. 

Simulation studies confirmed that maximum likelihood estimation in the zero-or-one inflated beta regression model is highly sensitive to outliers. The proposed robust estimators performed similarly to MLE in the absence of contamination but significantly outperformed it under contaminated data. The algorithm for selecting the tuning constant was effective, identifying the need for robustness and returning MLE when no contamination was present. It is important to note that the performance of this algorithm in selecting the tuning constant for the beta model has already been studied in \cite{ribeiro2023} (see also \cite{Maluf2024}). However, to date, we are unaware of any work that has implemented and studied the performance of a data-driven algorithm for selecting the tuning constant in binary data regression, as developed here. In addition, robust tests performed better than conventional ones, providing empirical levels close to nominal levels. The usual Wald test, however, may yield distorted results and is not recommended when there are outliers in the data.

The methodology used here can be applied for defining robust estimators for different classes of inflated models. Future works will address robust estimation in zero-and-one inflated beta regression models and the development of an R package for fitting robust estimators for inflated beta regression models.

\section*{Appendix}

We present the matrices $V_{\bm{\theta}, \alpha}(\bm{\upsilon})$ for the proposed robust estimators. For  M-LSE, we have 
$$V_{\bm{\theta}, \alpha}(\bm{\upsilon}) = \textbf{J}_\alpha^{-1}(\bm{\upsilon}) \textbf{K}_\alpha(\bm{\upsilon})\textbf{J}_\alpha^{-1}(\bm{\upsilon}),$$
where
\begin{align*}
\textbf{J}_\alpha(\bm{\upsilon}) = -
\begin{bmatrix}
(1-\alpha) \bm{\mathcal{X}}^{\top} \textbf{A} \textbf{B}_1 {\textbf{T}}_\mu^2 \bm{\Phi}^2 \textbf{V} \bm{\mathcal{X}} & \bm{\mathcal{X}}^{\top} \textbf{A} \textbf{B}_1 {\textbf{T}_\mu} \textbf{T}_\phi^{\ast} \textbf{C} \bm{\mathcal{Z}}\\
\bm{\mathcal{Z}}^{\top} \textbf{A}\textbf{B}_1 {\textbf{T}_\mu} \textbf{T}_\phi^{\ast} \textbf{C} \bm{\mathcal{X}} & (1-\alpha)^{-1}\bm{\mathcal{Z}}^{\top}\textbf{A} \textbf{B}_1 {\textbf{T}_\phi^{\ast}}^2 \textbf{D} \bm{\mathcal{Z}}
\end{bmatrix}
\end{align*}
and
\begin{align*}
\textbf{K}_\alpha(\bm{\upsilon}) = 
\begin{bmatrix}
\bm{\mathcal{X}}^{\top} \textbf{A}\textbf{B}_2 {\textbf{T}}_\mu^2 \bm{\Phi}^2 \textbf{V}_{1+\alpha} \bm{\mathcal{X}} & (1-\alpha)^{-1} \bm{\mathcal{X}}^{\top} \textbf{A}\textbf{B}_2 {\textbf{T}_\mu} \textbf{T}_\phi^{\ast} \textbf{C}_{1+\alpha} \bm{\mathcal{Z}}\\
(1-\alpha)^{-1}\bm{\mathcal{Z}}^{\top} \textbf{A}\textbf{B}_2 {\textbf{T}_\mu} \textbf{T}_\phi^{\ast} \textbf{C}_{1+\alpha} \bm{\mathcal{X}} & (1-\alpha)^{-2}\bm{\mathcal{Z}}^{\top} \textbf{A}\textbf{B}_2 {\textbf{T}_\phi^{\ast}}^2 \textbf{D}_{1+\alpha} \bm{\mathcal{Z}}
\end{bmatrix},
\end{align*}
in which $\textbf{A} = \diag\{ 1- \vartheta_i; i=1,\ldots, n \}$, $\textbf{B}_j = \diag\{b_{i, j}; i=1,\ldots, n \}$, $j=1,2$,
\[
b_{i, 1} = \dfrac{B(\mu_i \phi_i, (1-\mu_i) \phi_i)^{1-\alpha}}{B(\mu_{i} {\phi}_{i, 1-\alpha}, (1-\mu_{i}) {\phi}_{i, 1-\alpha})},
\quad b_{i, 2} = \dfrac{B(\mu_{i} {\phi}_{i, 1+\alpha}, (1-\mu_{i}) {\phi}_{i, 1+\alpha})}{B(\mu_i \phi_i, (1-\mu_i) \phi_i)^{2\alpha} B(\mu_{i} {\phi}_{i, 1-\alpha}, (1-\mu_{i}) {\phi}_{i,  1-\alpha})},
\]
$\textbf{T}_\mu = \diag\{t_{i, \mu}; i=1,\ldots, n \}$, $\textbf{T}_\phi^{\ast} = \diag\{t_{i, \phi}; i=1,\ldots, n \}$, $t_{i, \mu} = \left[ g'_\mu(\mu_{i}) \right]^{-1}$, $ t_{i, \phi} = \left[ g'_\phi({\phi}_{i, 1-\alpha}) \right]^{-1}$,
$\bm{\Phi} = \diag\{\phi_{i}; i=1,\ldots, n \}$, $\textbf{V} = \diag\{v_i; i=1,\ldots, n \}$, $v_i=\psi'(\mu_i \phi_i) + \psi'((1-\mu_i) \phi_i)$, $\textbf{V}_{1+\alpha} = \diag\{v_{i, 1+\alpha}; i=1,\ldots, n\}$, $v_{i, 1+\alpha}  = \psi'(\mu_{i} {\phi}_{i, 1+\alpha}) + \psi'((1-\mu_{i}) {\phi}_{i, 1+\alpha})$, $\textbf{C} = \diag\{c_i; i=1,\ldots, n \}$, $c_i  =  \phi_{i} \left[ \mu_{i}\psi'(\mu_i \phi_i) - (1-\mu_{i}) \psi'((1-\mu_i) \phi_i) \right]$, $\textbf{D} = \diag\{d_i; i=1,\ldots, n \}$, $\textbf{D}_{1+\alpha} = \diag\{d_{i, 1+\alpha}; i=1,\ldots, n \}$, $d_i  =   \mu_{i}^2 \psi'(\mu_i \phi_i) + (1-\mu_{i})^2 \psi'((1-\mu_i) \phi_i) -\psi'(\phi_i)$, $d_{i, 1+\alpha}  =  \mu_{i}^2\psi'(\mu_{i} {\phi}_{i, 1+\alpha}) + (1-\mu_{i})^2 \psi'((1-\mu_{i}) {\phi}_{i, 1+\alpha}) - \psi'( {\phi}_{i, 1+\alpha})$, $\textbf{C}_{1+\alpha} = \diag\{c_{i, 1+\alpha}; i=1,\ldots, n \}$,  $c_{i, 1+\alpha} = \phi_{i} [ \mu_{i}\psi'(\mu_{i} {\phi}_{i, 1+\alpha}) - (1-\mu_{i}) \psi'((1-\mu_{i}) {\phi}_{i, 1+\alpha})]$, $\bm{\mathcal{X}} = (\bm{\mathcal{X}}_1, \ldots, \bm{\mathcal{X}}_{n})^\top$, and $\bm{\mathcal{Z}} = (\bm{\mathcal{Z}}_1, \ldots, \bm{\mathcal{Z}}_{n})^\top$. 

For  M-LME, we have  
$$V_{\bm{\theta}, \alpha}(\bm{\upsilon}) = \bm{\Lambda}_\alpha^{-1}(\bm{\upsilon}) \bm{\Sigma}_\alpha(\bm{\upsilon})\bm{\Lambda}_\alpha^{-1}(\bm{\upsilon}),$$
where
\begin{align*}
\bm{\Lambda}_\alpha(\bm{\theta}) = 
\begin{bmatrix}
\bm{\mathcal{X}}^{\top} \textbf{A} \bm{\Lambda}_{11}^{(1+\alpha)} \bm{\mathcal{X}} & \bm{\mathcal{X}}^{\top}\textbf{A} \bm{\Lambda}_{12}^{(1+\alpha)} \bm{\mathcal{Z}}\\
\bm{\mathcal{Z}}^{\top}\textbf{A} \bm{\Lambda}_{12}^{(1+\alpha)} \bm{\mathcal{X}} & \bm{\mathcal{Z}}^{\top} \textbf{A}\bm{\Lambda}_{22}^{(1+\alpha)} \bm{\mathcal{Z}}
\end{bmatrix}
\end{align*}
and
\begin{align*}
\bm{\Omega}_\alpha(\bm{\theta}) = 
\begin{bmatrix}
\bm{\mathcal{X}}^{\top} \textbf{A} \left[ \bm{\Lambda}_{11}^{(1+2\alpha)} - {\bm{\Lambda}_{1}^{(1+\alpha)}}^2 \right] \bm{\mathcal{X}} & \bm{\mathcal{X}}^{\top}\textbf{A} \left[ \bm{\Lambda}_{12}^{(1+2\alpha)} - {\bm{\Lambda}_{1}^{(1+\alpha)} \bm{\Lambda}_{2}^{(1+\alpha)}} \right] \bm{\mathcal{Z}}\\
\bm{\mathcal{Z}}^{\top}\textbf{A} \left[ \bm{\Lambda}_{12}^{(1+2\alpha)} - {\bm{\Lambda}_{1}^{(1+\alpha)} \bm{\Lambda}_{2}^{(1+\alpha)}} \right] \bm{\mathcal{X}} & \bm{\mathcal{Z}}^{\top} \textbf{A} \left[ \bm{\Lambda}_{22}^{(1+2\alpha)} - {\bm{\Lambda}_{2}^{(1+\alpha)}}^2 \right] \bm{\mathcal{Z}}
\end{bmatrix},
\end{align*}
in which $\bm{\Lambda}_{j}^{(\alpha)} = \diag \{ \Lambda_{j, i}^{(\alpha)}; i=1, \ldots, n \}$, for $j=1, 2$, $\bm{\Lambda}_{11}^{(\alpha)} = \diag \{ \Lambda_{11, i}^{(\alpha)}; i=1, \ldots, n  \}$, $\bm{\Lambda}_{12}^{(\alpha)} = \diag \{ \Lambda_{12, i}^{(\alpha)}; i=1, \ldots, n  \}$, $\bm{\Lambda_{22}}^{(\alpha)} = \diag \{ \Lambda_{22, i}^{(\alpha)}; i=1, \ldots, n \}$, with
\begin{align*}
\Lambda_{11, i}^{(\alpha)} & = \dfrac{\phi_i^2 \mathcal{K}_{i, \alpha}(\bm{\theta}) }{[g'_\mu(\mu_i)]^2} \left[ v_{i, \alpha} + (\mu_{i,\alpha}^{ \star} -\mu^\star_{i} )^2 \right], \\
\Lambda_{12, i}^{(\alpha)} & =\dfrac{\phi_i \mathcal{K}_{i, \alpha}(\bm{\theta}) }{g'_\mu(\mu_i)g'_\phi(\phi_i)} \left\{ \mu_i \left[ v_{i, \alpha} + (\mu_{i, \alpha}^{ \star} -\mu^\star_{i} )^2 \right] -  \psi'((1-\mu_{i})\phi_{i, \alpha} ) + (\mu_{i, \alpha}^{ \star} -\mu^\star_{i})(\mu_{i, \alpha}^{ \dagger} -\mu^\dagger_{i}) \right\},\\
\Lambda_{22, i}^{(\alpha)} & = \dfrac{ \mathcal{K}_{i, \alpha}(\bm{\theta}) }{[g'_\phi(\phi_i)]^2}\left\{ \mu_i^2  \psi'(\mu_{i}\phi_{i, \alpha}) + (1-\mu_i)^2\psi'((1-\mu_{i})\phi_{i, \alpha} ) - \psi'(\phi_{i, \alpha} ) \right. \\
& \hspace*{2.3cm} + \left. \left[\mu_i(\mu_{i, \alpha}^{\prime \star} -\mu^\star_{i}) + (\mu_{i, \alpha}^{\prime \dagger} -\mu^\dagger_{i})\right]^2 \right\}.
\end{align*}

\section*{Acknowledgments}
This study was financed in part by the Coordena\c{c}\~ao de Aperfei\c{c}oamento de Pessoal de N\'ivel Superior - Brazil (CAPES) - Finance Code 001 and by the Conselho Nacional de Desenvolvimento Cient\'ifico e Tecnol\'ogico - Brazil (CNPq). 


\end{document}